\documentclass[times,twocolumn,final,authoryear,dvipdfmx]{elsarticle} 

\usepackage{jasr}
\usepackage{framed,multirow}

\usepackage{amssymb}
\usepackage{latexsym}

\usepackage{graphics}

\usepackage{url}
\usepackage{xcolor}
\definecolor{newcolor}{rgb}{.8,.349,.1}

\usepackage[citebordercolor=white]{hyperref}

\journal{Advances in Space Research}

\begin{document}

\verso{Shin Toriumi}

\begin{frontmatter}

\title{Flux emergence and generation of flare-productive active regions}%

\author[1]{Shin \snm{Toriumi}\corref{cor1}}
\cortext[cor1]{Corresponding author: 
  Tel.: +81-70-1170-2784;
  fax: +81-42-759-8526;}
\ead{toriumi.shin@jaxa.jp}

\address[1]{Institute of Space and Astronautical Science, Japan Aerospace Exploration Agency, 3-1-1 Yoshinodai, Chuo-ku, Sagamihara, Kanagawa 252-5210, Japan}


\received{}
\finalform{}
\accepted{}
\availableonline{}
\communicated{}

\begin{abstract}
Solar flares and coronal mass ejections are among the most prominent manifestations of the magnetic activity of the Sun. The strongest events of them tend to occur in active regions (ARs) that are large, complex, and dynamically evolving. However, it is not clear what the key observational features of such ARs are, and how these features are produced. This article answers these fundamental questions based on morphological and magnetic characteristics of flare-productive ARs and their evolutionary processes, i.e., large-scale flux emergence and subsequent AR formation, which have been revealed in observational and theoretical studies. We also present the latest modeling of flare-productive ARs achieved using the most realistic flux emergence simulations in a very deep computational domain. Finally, this review discusses the future perspective pertaining to relationships of flaring solar ARs with the global-scale dynamo and stellar superflares.
\end{abstract}

\begin{keyword}
\KWD Solar flares\sep Coronal mass ejections\sep Sunspots\sep Active regions\sep Magnetohydrodynamics
\end{keyword}

\end{frontmatter}


\section{Introduction}\label{sec:intro}

The Sun ceaselessly exhibits activity events of various scales over the whole spectrum of electromagnetic waves. Solar flares and coronal mass ejections (CMEs) are among the most prominent energy-releasing phenomena in the current solar system \citep{2011SSRv..159...19F,2011LRSP....8....6S}. Most powerful events often emanate from active regions (ARs) that harbor substantial magnetic non-potentiality.

ARs are created owing to the emergence of dynamo-generated toroidal magnetic flux from the subsurface layer (flux emergence; \citealt{1955ApJ...121..491P}). In the convection zone, where the gas pressure surpasses the magnetic pressure (i.e., plasma-$\beta$ is greater than unity), magnetic flux is buffeted by convective turbulence and accumulates twist during its ascent. However, as the flux reaches the photosphere and moves further into the atmosphere, where the plasma density drastically decreases and the magnetic field is dominant ($\beta<1$), the flux begins to build up ARs. Owing to the drastic reduction of the external gas pressure, the flux expands into the atmosphere, and the accumulated twist can be released. If an AR has a sufficient amount of magnetic flux, sunspots are gradually produced, and flares and CMEs may occasionally occur if the free magnetic energy ($\Delta E_{\rm mag}$) stored in the corona is suddenly released, where
\begin{eqnarray}
  \Delta E_{\rm mag}=\int\frac{B^{2}}{8\pi}\,dV - \int\frac{B_{\rm pot}^{2}}{8\pi}\,dV
\end{eqnarray}
and $\mbox{\boldmath $B$}$ and $\mbox{\boldmath $B$}_{\rm pot}$ denote the actual and potential magnetic fields, respectively.

\begin{figure}
\centering
\includegraphics[width=0.45\textwidth]{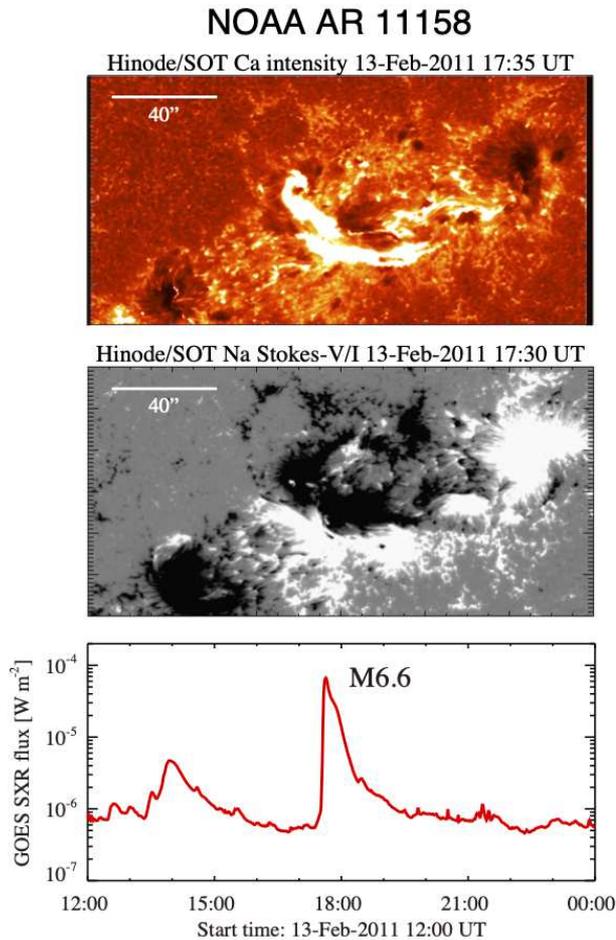}
\caption{M6.6-class flare in NOAA AR 11158 as observed by Hinode/SOT \citep{2007SoPh..243....3K,2008SoPh..249..167T} with the GOES soft X-ray light curve. The detailed analysis suggests that the hook-shaped magnetic field on the PIL triggered the event \citep{2012ApJ...760...31K,2013ApJ...778...48B}. Figure reproduced from \citet{2013ApJ...773..128T} by permission of the AAS.}
\label{fig:11158}
\end{figure}

Fig. \ref{fig:11158} presents the observation of an M-class flare (occurred in February 2011) in NOAA AR 11158. The chromospheric image (top panel) displays two bright patches, known as flare ribbons, that extend at both sides of the polarity inversion line (PIL), where the magnetic polarity ($B_{z}$, or the Stokes-I/V signal in the middle panel) alters its sign.

In the flare theories such as the CSHKP model \citep[after][]{1964NASSP..50..451C,1966Natur.211..695S,1974SoPh...34..323H,1976SoPh...50...85K} and its variations, the flare ribbons brighten in response to the injection of energy that precipitates along the loops, sourcing from the site of magnetic reconnection in the corona. The heated chromospheric plasma {\it evaporates} and fills the coronal loops, which leads to the stark enhancement of the soft X-ray flux, as shown in the bottom panel of Fig. \ref{fig:11158}. The loops gradually cool down with time through radiation and thermal conduction, and eventually, the X-ray light curve returns to the background level.

As a result of reconnection, a helical flux rope is ejected upward, probably with the support of plasma instabilities, and, if successful, it develops into a CME that travels through the interplanetary space. The flare event in Fig. \ref{fig:11158} was reportedly accompanied by the CME eruption \citep[e.g.,][]{2012ApJ...745L...4L,2018NatCo...9..174I}.

The overall description above, i.e., from the dynamo, flux emergence, AR formation, to the subsequent occurrence of flares and CMEs, is accepted in the research community. However, in this article, we address the following two key questions:
\begin{itemize}
\item What are the important observational features of the flare-productive ARs?
\item If there exist any, how are these features created?
\end{itemize}
As the genesis of flaring ARs is related to the prediction and forecasting of flare eruptions, it is recognized as one of the primary targets in solar physics to be explored in the next decade \citep[see, e.g.,][]{2015AdSpR..55.2745S}.

In the rest of this article, we focus on the flux emergence and formation of ARs that are prone to major solar flares, and thus, for the understanding of flares and CMEs, readers are advised to refer to other excellent reviews \citep[e.g.,][and references therein]{2002A&ARv..10..313P,2011LRSP....8....1C,2011SSRv..159...19F,2011LRSP....8....6S,2015SoPh..290.3425J,2017LRSP...14....2B}.

Section \ref{sec:obs} provides an overview of flare-productive ARs by reviewing the observational features of the ARs. Section \ref{sec:sim} presents theoretical studies to numerically reproduce the formation processes of flaring ARs. Section \ref{sec:r2d2} introduces the latest modeling attempts of flaring ARs with the most realistic setup. Section \ref{sec:summary} summarizes the paper and discusses some remaining problems.

\section{Observational aspects}\label{sec:obs}

This section deals with the observational aspects of flare-productive ARs. Since the discovery of solar flares \citep{1859MNRAS..20...13C,1859MNRAS..20...15H}, significant effort has been made to scrutinize the mechanism and origin of eruptions. Considering that strong flares mostly occur in ARs, we can directly examine the relationship between the flares and morphological and magnetic characteristics of their hosting sunspots, which we elaborate in the following subsections.

\subsection{$\delta$-sunspots and flare activity}\label{sec:obs:delta}

Observers have noticed that the structure of sunspots is one of the key factors that determine the flare activity. For instance, in the following Mt. Wilson sunspot classification, $\delta$-spots are known to be very flare-active \citep{1960AN....285..271K}:
\begin{itemize}
\item $\alpha$, a unipolar spot group
\item $\beta$, a single bipolar spot group of both positive and negative polarities
\item $\gamma$, a complex spot group in which spots of both polarities are distributed so irregularly as to prevent classification as a $\beta$ group
\item $\delta$, a spot group in which umbrae of opposite polarities are separated by less than 2$^{\circ}$ and situated within the common penumbra
\end{itemize}
\citet{2000ApJ...540..583S} showed that the flare magnitude increases with the size and complexity of the spot groups. According to these authors, all $\ge$X4-class flares occur in spot groups of areas greater than 1000 MSH and classified as the most complex $\beta\gamma\delta$.

Major spaceweather events in history were mostly caused by $\delta$-spots. For instance, the great geomagnetic storm in March 1989, which caused a severe power outage in Quebec, Canada, and damaged devices at power plants in New Jersey, USA, is related to NOAA AR 5395 and its series of strong flares \citep[e.g.,][]{1989EOSTr..70.1479A}. The images and magnetograms of this AR, as shown in \citet{1991ApJ...380..282W}, reveal the existence of large complex-shaped $\delta$-spots.

\citet{1987SoPh..113..267Z} categorized the formation of $\delta$-spots into three types:
\begin{itemize}
\item Type 1: spots emerging all at once, intertwined
\item Type 2: satellite spots near larger older spots
\item Type 3: collision of two bipoles
\end{itemize}
The representative spot group of Type 1 is NOAA 5395 (March 1989), but other events, such as NOAA 5629 (August 1989), 5747 (October 1989), and 6659 (June 1991), have been extensively analyzed \citep{1993SoPh..143..107T,1994SoPh..150..199S}. A clear example of Type 2 is NOAA 10930 (December 2006). An X3.4-class flare occurred between sunspots of positive and negative polarities, but the positive spot was a part of a newly-emerging bipole, whose counterpart was quickly disappearing \citep{2007PASJ...59S.779K,2008ApJ...687..658W}: Observations of NOAA 10930 are summarized in Section 8 of \citet{2019PASJ...71R...1H}. NOAA 11158 (Fig. \ref{fig:11158}: February 2011) is the typical Type 3 $\delta$-spot. In this event, two bipoles emerged and underwent fly-by motion (near head-on collision). A series of strong flares occurred between the positive and negative spots, originated from different emerging bipoles, at the center \citep{2013ApJ...764L...3C,2014SoPh..289.3351T}.

\begin{figure*}
\centering
\includegraphics[width=0.9\textwidth]{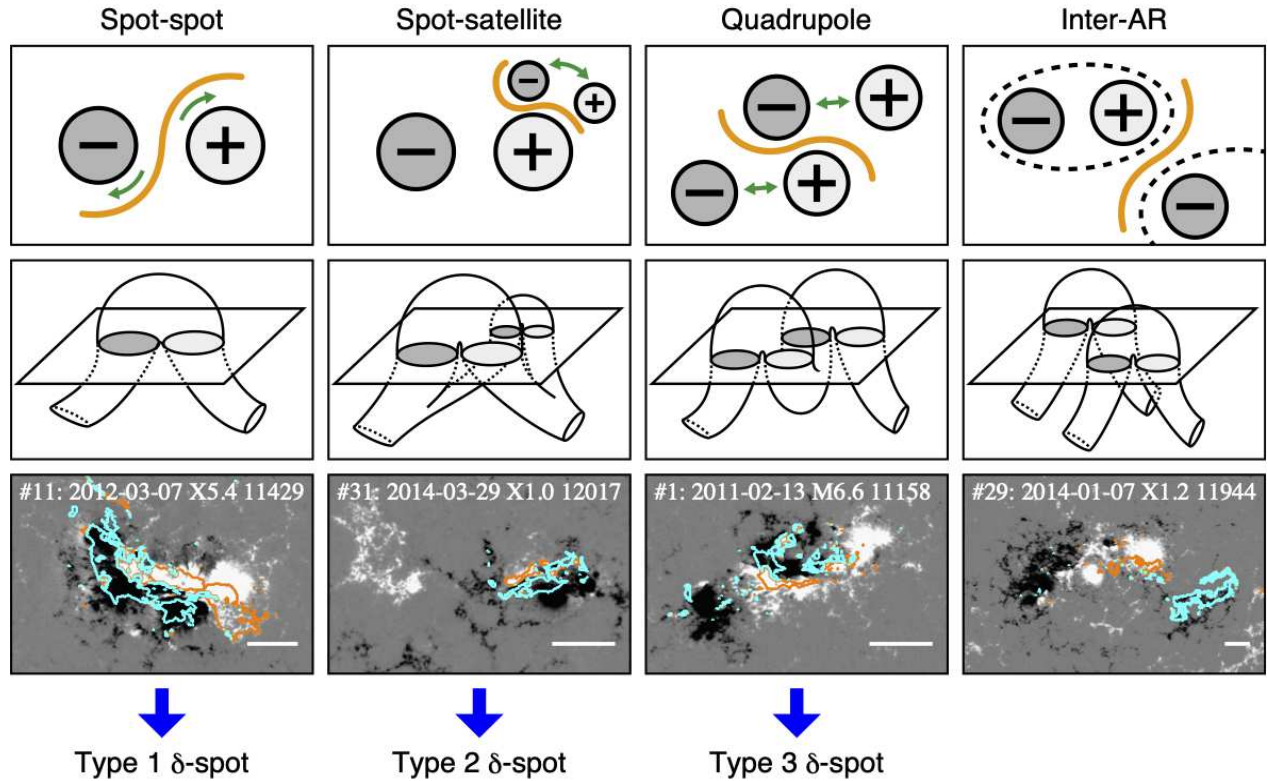}
\caption{Four representative categories of flaring ARs. From top to bottom, the polarity distribution, possible 3D structures of magnetic fields, and the sample events. In the magnetograms (bottom panels), temporally stacked flare ribbons are indicated by orange and turquoise contours. Figures reproduced from \citet{2017ApJ...834...56T} by permission of the AAS.}
\label{fig:spot}
\end{figure*}

Fig. \ref{fig:spot} displays four categories of flaring ARs \citep{2017ApJ...834...56T}. Surveying 51 $\ge$M5-class events that occurred within 45$^{\circ}$ from the disk center over six years, these authors classified ARs based on their evolutionary patterns: (1) spot-spot, a complex, compact $\delta$-spots, in which an elongated PIL extends across the entire AR (probably equivalent to Type 1 $\delta$-spot); (2) spot-satellite, where a new emerging flux collides with an existing spot (Type 2); and (3) quadrupolar, the collision of two bipoles (Type 3). It was found that 83\% of the analyzed ARs displayed $\delta$-structures. However, two flares (X1.2 and M5.1) occurred at the boundaries among separated independent ARs (inter-AR cases), indicating that X-class events do not require $\delta$-spots.

Therefore, it is possible to categorize the formation of flaring ARs based on the apparent motions of individual polarities. However, we cannot investigate the subsurface flux emergence from direct optical observations. This is why the series of flux emergence simulations have been conducted.

\subsection{Observational characteristics}

In flaring ARs, PILs show a strong horizontal field ($>$4000 G), strong $B_{z}$ gradient ($\sim$100 G Mm$^{-1}$), and strong magnetic shear (80$^{\circ}$--90$^{\circ}$ from the direction of the potential field) \citep{1958IzKry..20...22S,1984SoPh...91..115H,1996ApJ...456..861W,2007ApJ...655L.117S}. Recently, \citet{2018ApJ...852L..16O}, \citet{2018RNAAS...2....8W}, and \citet{2020ApJ...895..129C} detected superstrong magnetic fields of $\gtrsim$6000 G in the transverse fields of $\delta$-spots light bridges (i.e., PILs). These values are probably the strongest ever measured on the Sun, including the sunspot umbrae. These strong-field, strong-gradient, highly-sheared PILs can be direct manifestation of the non-potentiality of magnetic fields. Therefore, observational parameters that characterize flaring PILs have been extensively used to predict flares and CMEs \citep[e.g.,][]{2002ApJ...569.1016F,2003ApJ...595.1277L,2007ApJ...655L.117S,2020Sci...369..587K}.

Such flaring PILs are often produced in the course of spot rotations and counter-streaming flows at PILs \citep{1976SoPh...47..233H}. For instance, the flares in NOAA 10930 occurred along the PIL between the two spots, and \citet{2009SoPh..258..203M} reported a fast spot rotation with an angular speed of $\sim 8^{\circ}$ hr$^{-1}$ before the X-class event, which sheared the magnetic fields around the PIL and supplied free energy. \citet{2019ApJ...871...67C} pointed out that flaring PILs are created by the relative movement of the spots, especially when the shearing and flux cancellation develop owing to the spot collision (collisional shearing). These PILs sometimes exhibit the ``magnetic channel,'' an alternating pattern of elongated positive and negative polarities \citep{1993Natur.363..426Z,1993SoPh..144...37Z}.

The complexity of ARs are often quantified by the magnetic helicity, a measure of magnetic topology such as twists, kinks, and internal linkage \citep{1956RvMP...28..135E}. The helicity of the magnetic field $\mbox{\boldmath $B$}$ fully contained in volume $V$ is defined as
\begin{eqnarray}
  H=\int_{V} \mbox{\boldmath $A$}\cdot\mbox{\boldmath $B$}\, dV,
\end{eqnarray}
where $\mbox{\boldmath $A$}$ is the vector potential of $\mbox{\boldmath $B$}$, i.e., $\mbox{\boldmath $B$}=\nabla\times\mbox{\boldmath $A$}$. In practical situations in solar physics, the magnetic field lines cross the boundary (e.g., the photosphere); therefore, it is more convenient to use the relative helicity,
\begin{eqnarray}
  H_{\rm R}=\int_{V} (\mbox{\boldmath $A$}+\mbox{\boldmath $A$}_{0})\cdot(\mbox{\boldmath $B$}-\mbox{\boldmath $B$}_{0})\, dV,
\end{eqnarray}
where $\mbox{\boldmath $A$}_{0}$ and $\mbox{\boldmath $B$}_{0}$ are the reference values of $\mbox{\boldmath $A$}$ and $\mbox{\boldmath $B$}$, respectively \citep{1984JFM...147..133B,finn1985}. $H_{R}$ has the advantage of being gauge invariant, and, in many cases, the potential field $\mbox{\boldmath $B$}_{\rm pot}\,(=\nabla\times\mbox{\boldmath $A$}_{\rm pot})$ is chosen as the reference field. Numerous observations have claimed the relationship between helicity injection and flare occurrence \citep[see, e.g.,][and references therein]{2009AdSpR..43.1013D,2016SSRv..201..147V}.

The ``magnetic tongues,'' the two extended polarities on magnetograms at both sides of the PILs, especially when they have a yin-yang pattern, are assumed to be the vertical projection of the poloidal component of the twisted emerging magnetic flux \citep{2000ApJ...544..540L}. Therefore, the arrangement of the tongues and PILs represent the sign of helicity in ARs, and observations reveal that flaring ARs show such ying-yang tongues \citep[e.g.,][]{2014SoPh..289.2041M,2015SoPh..290..727P}.

In the atmosphere above the PIL, magnetic flux ropes (bundles of twisted field lines) are sometimes observed as ``sigmoids'' in soft X-rays \citep{1996ApJ...464L.199R,1999GeoRL..26..627C} and ``filaments'' in H$\alpha$ images \citep{2002SoPh..207..111P}. The observations indicate that the filament formation is in favor of the model proposed by \citet{1989ApJ...343..971V}, in which converging and shearing flows around the PIL cause cancellation and twisting of the field lines to eventually produce a flux rope (filaments/prominences) above the PIL.

Other characteristics, such as non-neutralized electric currents, structural complexity of photospheric magnetograms, non-thermal broadening of coronal emission lines, and helioseismic (subsurface) signatures, exist. Moreover, the triggering of flare eruptions are often associated with the rapid development of photospheric magnetic fields. For more comprehensive discussion, readers may refer to Section 3 of \citet{2019LRSP...16....3T}.

\section{Theoretical aspects}\label{sec:sim}

As ARs are created through flux emergence from below the solar surface, it is essential to trace the evolution of magnetic flux from the deeper convection zone into the atmosphere. To realize this, the magnetohydrodynamic (MHD) simulations of flux emergence have been performed \citep{2008JGRA..113.3S04A,2009LRSP....6....4F,2014LRSP...11....3C,2014SSRv..186..227S}. In typical flux emergence simulations, in the stratified atmospheric layers representing the solar convection zone, photosphere, chromosphere, and corona, a flux tube is initially embedded into the convection zone. The flux tube starts rising by, for instance, perturbing velocity fields inside the tube or partially reducing the density from the tube to trigger the buoyant emergence (Parker instability: \citealt{1979cmft.book.....P}). As it ascents to the atmosphere, the flux tube builds up bipolar spots in the photosphere and magnetic loops in the corona.

To explain the generation of flare-productive ARs, several scenarios have been suggested and tested by performing such flux emergence simulations. In this section, we review the past attempts to theoretically model the observed characteristics of flaring ARs.

\subsection{Idealized models}\label{sec:sim:ideal}

 Many of the flux emergence simulations conducted so far fall into the category of idealized models, where the subsurface layer is given as an adiabatically stratified, plane parallel atmosphere; i.e., the interior is not convectively unstable. This treatment sets aside the complex effect of turbulent convection and allows modelers to focus on the essential physical processes.

\begin{figure*}
\centering
\includegraphics[width=0.9\textwidth]{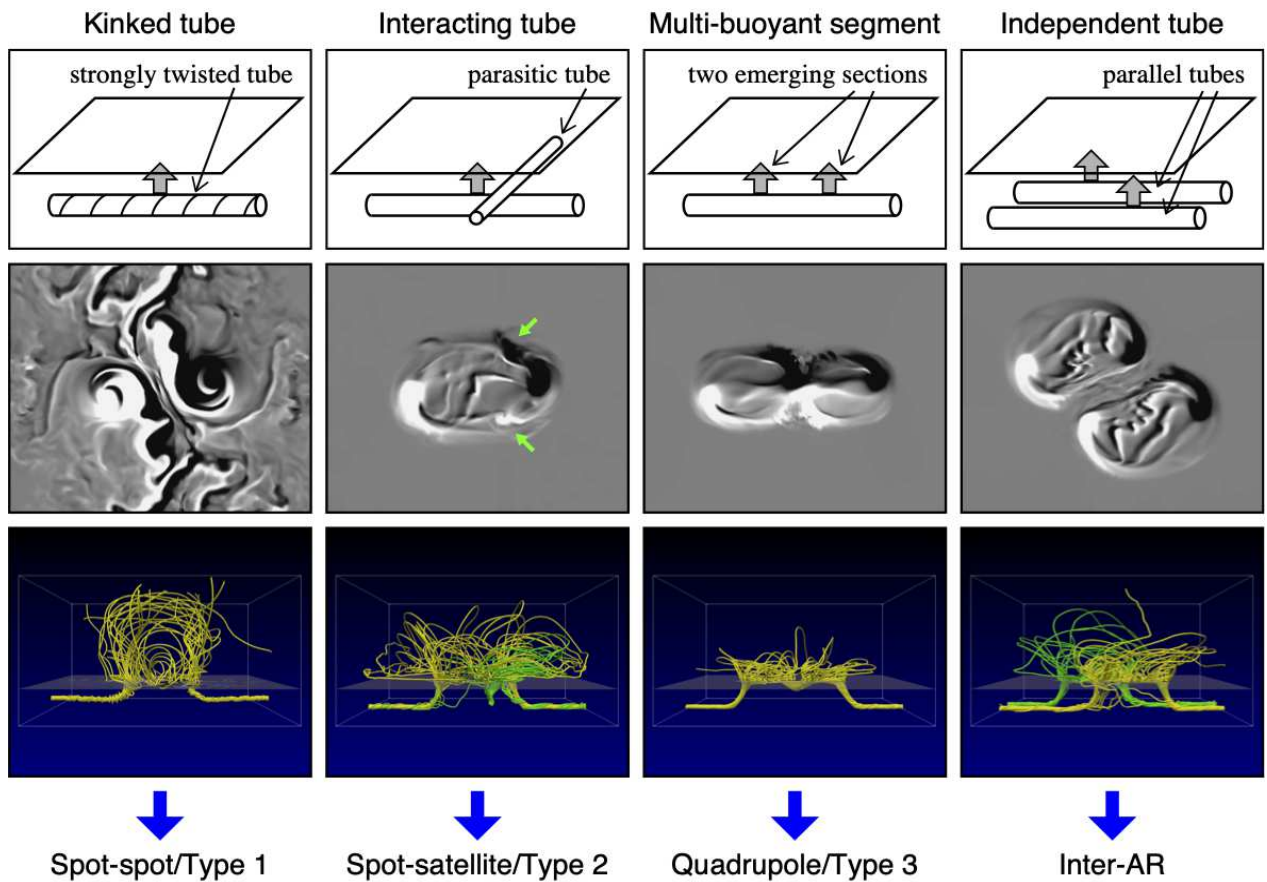}
\caption{Four representative flux emergence models of flare-productive ARs. The green arrows in the second column point to the satellite spots, which originate from the parasitic flux tube. Figures reproduced from \citet{2017ApJ...850...39T} by permission of the AAS.}
\label{fig:model}
\end{figure*}

One of the representative scenarios of $\delta$-spot formation is the ``kinked tube'' model, in which a strongly twisted tube rises owing to the kink instability \citep{1991SoPh..136..133T}. Using the anelastic MHD code, \citet{1999ApJ...521..460F} calculated the emergence of a kink-unstable flux tube in the interior and discovered that the horizontal slice of the emerging flux tube displays a compact bipolar pair with sheared PIL. Although this morphology is reminiscent of $\delta$-spots, these simulations were limited to the convection zone due to the anelastic approximation. \citet{2015ApJ...813..112T} coped with this problem by solving fully compressible MHD equations and extending the computational box to the corona. Interestingly, their modeled spots present, in spite of the simple bipolar configuration, a quadrupolar structure comprising main bipolar spots with an elongated pair in between (left column of Fig. \ref{fig:model}). Owing to the release of the helicity, the main spots show strong rotational motions, which is one of the key characteristics of the flaring ARs. The complex morphology in the magnetogram suggests that the kinked tube model may explain Type 1 $\delta$-spots \citep[see also][]{2018ApJ...864...89K}.

The second group is the ``interacting tube'' model, in which multiple flux bundles interfere with each other to build a single, perhaps complex, AR \citep[e.g.,][]{2007A&A...470..709M}. As shown in Fig. \ref{fig:model}, second column from the left, a parasitic flux tube is placed right above the main tube along the orthogonal direction. Lifted by the main tube, the parasitic tube produces satellite spots in the proximity to the main spot (indicated by green arrows). Therefore, this scenario may be one of the mechanisms of Type 2 $\delta$-spots, in which newly emerging flux appears within the penumbra of the existing spot \citep[see also][]{2018ApJ...857...83J}.

In the ``multi-buoyant segment'' scenario, a single flux tube becomes buoyant at two segments along the tube, probably caused by the convective pinning at the middle, and produces a pair of emerging bipoles on the surface, i.e., a quadrupolar AR \citep{2014SoPh..289.3351T}. In this case, the two bipoles cause head-on collision in between, and as a result, strongly packed positive and negative polarities are created with a highly sheared PIL in the center of the quadrupolar AR (Fig. \ref{fig:model}, third column). The overall evolution resembles Type 3 $\delta$-spots \citep{2015ApJ...806...79F,2019A&A...630A.134S}.

These models succeeded in reproducing $\delta$-configurations, sheared PILs, flux ropes, and others, all of which are the essential ingredients of the flaring ARs. \citet{2017ApJ...850...39T} performed a systematic survey of the aforementioned models using similar numerical conditions with as little difference as possible (Fig. \ref{fig:model}). Key findings are summarized below.
\begin{itemize}
\item Highly sheared PILs are formed owing to the combined effects of advection, stretching, and compression of magnetic fields, exerted by the spot motion.
\item Flux ropes are created above the PIL. However, access to the outer atmosphere (interplanetary space) depends on the magnetic structure of the entire AR, and this may determine the CME productivity.
\item Free energy is accumulated in the corona in the form of current sheets. Magnetic parameters that predict solar flares with high accuracy \citep[e.g.,][]{2015ApJ...798..135B} reflect the free magnetic energy stored in the corona well.
\end{itemize}

\subsection{Realistic models}

Contrary to idealized models, flux emergence models that incorporate the effect of thermal convection enable us to understand complex realistic aspects of the evolution, in particular, the coupling of magnetic field and turbulence. For instance, \citet{2010ApJ...720..233C} modeled the emergence of a twisted flux tube from the {\it convective} convection zone and eventual formation of proto-sunspots. By kinematically inserting the flux tube from the bottom boundary, placed at 7.5 Mm below the photosphere, as a sequentially updated bottom boundary, the authors succeeded in reproducing spots with umbral dots, light bridges, and penumbral filaments, which are the manifestation of magnetoconvection \citep[see also][]{2014ApJ...785...90R}.

Using the updated version of the same code, which is now extended to the solar corona, \citet{2019NatAs...3..160C} mimicked the emergence of small satellite spots within the penumbra of the existing spot to reproduce X-class flares in NOAA 12017 (April 2014). Their numerical results explain the properties of solar flares, such as temporal evolution of X-ray light curve, chromospheric evaporation, and flare ribbon expansion.

The interplay between the global-scale convection and magnetic flux has been strenuously investigated by using the 3D spherical shell models. For example, \citet{2013ApJ...762...73N} simulated the self-consistent generation of $\Omega$-loops from a bundle of toroidal flux, whereas \citet{2013ApJ...762....4J} modeled the emergence of a magnetic flux tube that is placed at the depth of the convection zone. \citet{2018ApJ...857...83J} also investigated the interaction of multiple emerging flux tubes and pointed out the possibility that such interactions introduce complexity to ARs. Although these computations are limited to the convection zone due to the anelastic approximation and, thus, cannot simulate the emergence into the photosphere, they revealed the importance of the turbulent background convection and the mutual interaction of flux tubes.

\section{New twist}\label{sec:r2d2}

Flux emergence simulations have succeeded in explaining various aspects of the mechanisms behind the formation of flare-producing ARs. However, most simulations are idealized: many models lack the realistic convection zone, or the emergence is triggered by artificially imposed buoyancy. Even in convective flux emergence simulations, the bottom boundary is not deep enough, at most $-30$ Mm \citep{2017ApJ...846..149C}, and flux tubes are kinematically injected from the bottom boundary. However, in reality, ARs are formed as the outcome of interaction between magnetic flux and background turbulent convection. To elucidate the effect and importance of deep large cells on the flux emergence, there has been a demand for detailed calculations.

\begin{figure}
\centering
\includegraphics[width=0.48\textwidth]{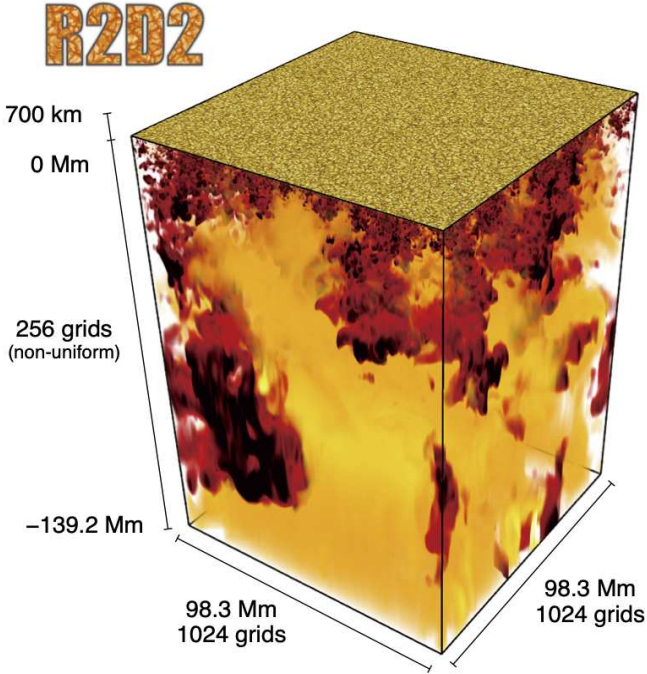}
\caption{Volume rendering of entropy perturbation in the {\it R2D2} simulation that is normalized by the averaged entropy at each depth level. Figure courtesy of H. Hotta.}
\label{fig:r2d2}
\end{figure}

To overcome these issues, \citet{2019ApJ...886L..21T} leveraged the newly developed radiative MHD code {\it R2D2} \citep{2019SciA....5.2307H,2020MNRAS.494.2523H}. In brief, this code reproduces the realistic solar convection in a very deep domain by employing the reduced sound-speed technique \citep{2005ApJ...622.1320R,2012A&A...539A..30H} and allows for the exploration of magnetoconvective evolution in the whole solar convection zone. First, the 3D computation box that spanned from $-139.2$ Mm to $+700$ km in height with a horizontal width of $98.3\ {\rm Mm}\,\times 98.3\ {\rm Mm}$ was prepared, and the box was numerically resolved by a $1024\times 1024\times 256$ grid (see Fig. \ref{fig:r2d2}). Considering that the previous convective emergence models reached down to approximately $-30\ {\rm Mm}$ and that the solar convection zone is a strongly stratified medium, the simulation with the bottom boundary at $-140\ {\rm Mm}$ is a significant advancement: The gas pressure at $-140\ {\rm Mm}$ is eight orders of magnitude greater than that in the photosphere. The periodic boundary condition was applied to both horizontal directions.

After the development of thermal convection in the box, a twisted force-free flux tube was placed at the depth of $16.7\ {\rm Mm}$, given as the Lundquist field \citep{1951PhRv...83..307L},
\begin{eqnarray}
  B_{x}=B_{\rm tb}J_{0}(\alpha r), B_{\phi}=B_{\rm tb}J_{1}(\alpha r),
\end{eqnarray}
where $r$ is the radial distance from the tube axis, $B_{\rm tb}=10\ {\rm kG}$ the axial field strength, $J_{0}$ and $J_{1}$ the Bessel functions, $\alpha=a_{0}/R_{\rm tb}$, $a_{0}=2.404825$, and $R_{\rm tb}=7\ {\rm Mm}$ the tube's radius. Unlike previous models, the flux tube was neither endowed with an artificial buoyancy nor forcibly injected into the domain. The tube was simply advected by the turbulent background convection.

Note that using $16.7\ {\rm Mm}$ as the initial depth allows us to investigate the emergence from the critical layer, where the approximations assumed in the previous simulations become invalid (around $-20\ {\rm Mm}$ for the anelastic and thin flux tube approximations; \citealt{2009LRSP....6....4F}). The artificial influences caused by adding a flux tube into the domain at one time step where there was none at the previous step \citep[e.g.,][]{2008ApJ...687.1373C}, such as a lack of the perfect mechanical balance, can be circumvented using the flux injection method. However, these influences would be minor because vigorous turbulence drastically reforms the magnetic configuration (see the following paragraphs).

\begin{figure*}
\centering
\includegraphics[width=0.9\textwidth]{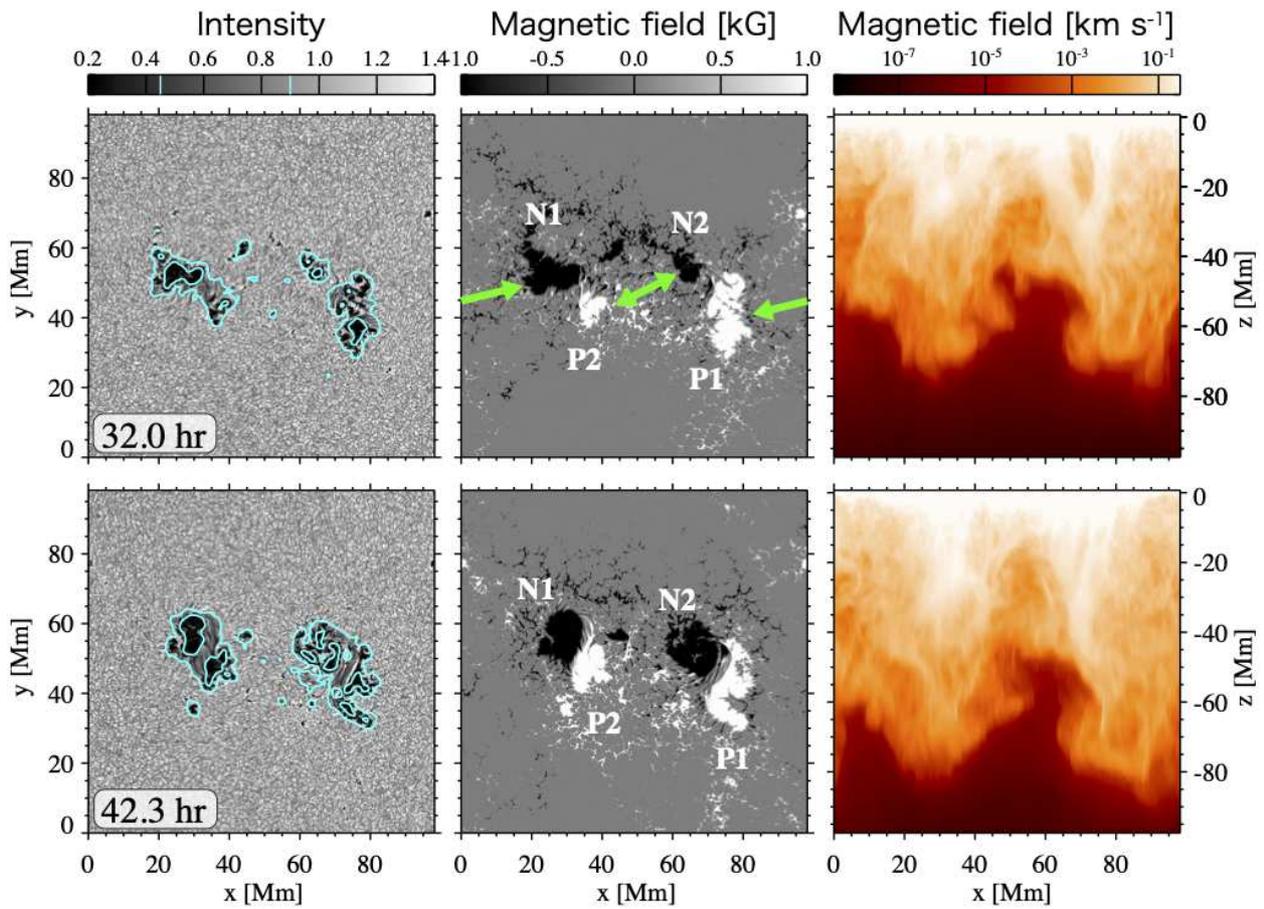}
\caption{Generation of $\delta$-spots in the {\it R2D2} simulation. From left to right, normalized emergent intensity, magnetic field strength $B_{z}$ (magnetogram), and normalized field strength $|\mbox{\boldmath $B$}|/(4\pi \rho_{0})^{1/2}$ (thus in units of velocity) on the vertical cross-section at 32.0 hr (upper row) and 42.3 hr (lower row) after the flux tube is initially placed. A pair of bipoles (P1-N1 and P2-N2) collide to produce $\delta$-spots N1-P2 and N2-P1. Figure reproduced from \citet{2019ApJ...886L..21T} by permission of the AAS.}
\label{fig:evolution}
\end{figure*}

Fig. \ref{fig:evolution} shows the evolutions of intensity, surface magnetogram, and magnetic field in the vertical cross-section. First, a large convection upflow elevates the flux tube at the side edge of the box at $x=0$ and $98.3\ {\rm Mm}$ (because of the periodic boundary condition), and a bipolar pair P1-N1 appears in the photosphere. As time progresses, the secondary pair P2-N2 emerges in the middle of the domain owing to another upflowing region and, consequently, a quadrupolar AR is formed. The right column demonstrates that two strong downdrafts (approximately $x=30$ and $70\ {\rm Mm}$) drag down the flux tube to make a W-shaped configuration. On the surface, the two emerging bipoles approach each other as they expand, and eventually, two $\delta$-spots, N1-P2 and N2-P1, are formed. The visibility in the intensity map shows that in each spot, the umbrae of opposite polarities reside in a common penumbra, satisfying the definition of the $\delta$-spot (Section \ref{sec:obs:delta}).

The PIL between the two spots within each $\delta$-spot shows a narrow lane of elongated granular cells (light bridge), and the polarity pattern resembles the magnetic channel structure. The previous simulations revealed that the light bridges in regular spots comprise granules, which, especially in the case of narrow light bridges, resemble the penumbra with an elongated magnetic field \citep{2010ApJ...720..233C,2015ApJ...811..138T,2015ApJ...811..137T}. The $\delta$-spot light bridge reproduced here shares a similar magnetoconvective structure.

In the photosphere, all four spots show counter-clockwise rotations owing to the releasing of the twist of the tube, which was imposed as the initial condition. As a result, a sheared (or counter-streaming) flow is excited around the PIL between the two neighboring spots of each $\delta$-spot, and this velocity field arranges the transverse component of the magnetic field to make the PIL highly sheared. The field strength pertaining to the PIL area far exceeds the equipartition value, which is of the order of 1 kG. \citet{2020MNRAS.498.2925H} performed high-resolution calculations, where the grid spacing is up to four times finer, and discovered that a very strong horizontal field of $\gtrsim$6000 G is formed at the edge of the $\delta$-spot light bridge. The field strength matches the recent observations made by \citet{2018ApJ...852L..16O} and others.

\begin{figure*}
\centering
\includegraphics[width=0.8\textwidth]{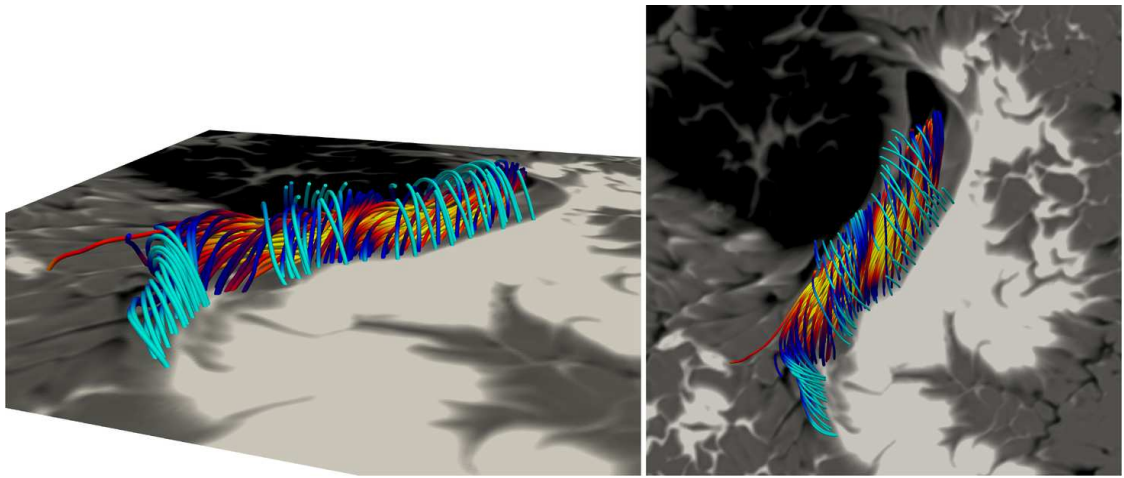}
\caption{Bird's eye view and top-down view of magnetic field lines above the $\delta$-spot N2-P1 in Fig. \ref{fig:evolution}. A helical flux rope structure is formed above the PIL. It should be noted that this flux rope is created as a result of magnetic reconnection between the field lines originating from the two independent but neighboring sunspots N2 and P1, rather than the direct emergence of the original flux tube. Figure reproduced from \citet{2019ApJ...886L..21T} by permission of the AAS.}
\label{fig:fluxrope}
\end{figure*}

Fig. \ref{fig:fluxrope} shows the field line structure above one of the $\delta$-spots in Fig. \ref{fig:evolution}. Driven by the counter-streaming flow in the photosphere, the overlying arcade field is sheared and forms a flux rope structure. The 6-kG horizontal field \citep{2020MNRAS.498.2925H} is created by the strong shear at the $\delta$-spot PIL, which is sustained owing to the downward magnetic force caused by the overlying arcade.

The above simulations demonstrate that the flux emergence in the deep-enough computational box self-consistently reproduces $\delta$-spots and other key features. The entire evolution of $\delta$-spots follows the multi-buoyant segment scenario, in which a single flux system in the interior rises at two locations to produce colliding bipoles. It should be emphasized that the $\delta$-spot formation is a result of coupling between the magnetic flux and turbulent background convection.

\section{Summary and discussion}\label{sec:summary}

\begin{figure*}
\centering
\includegraphics[width=0.9\textwidth]{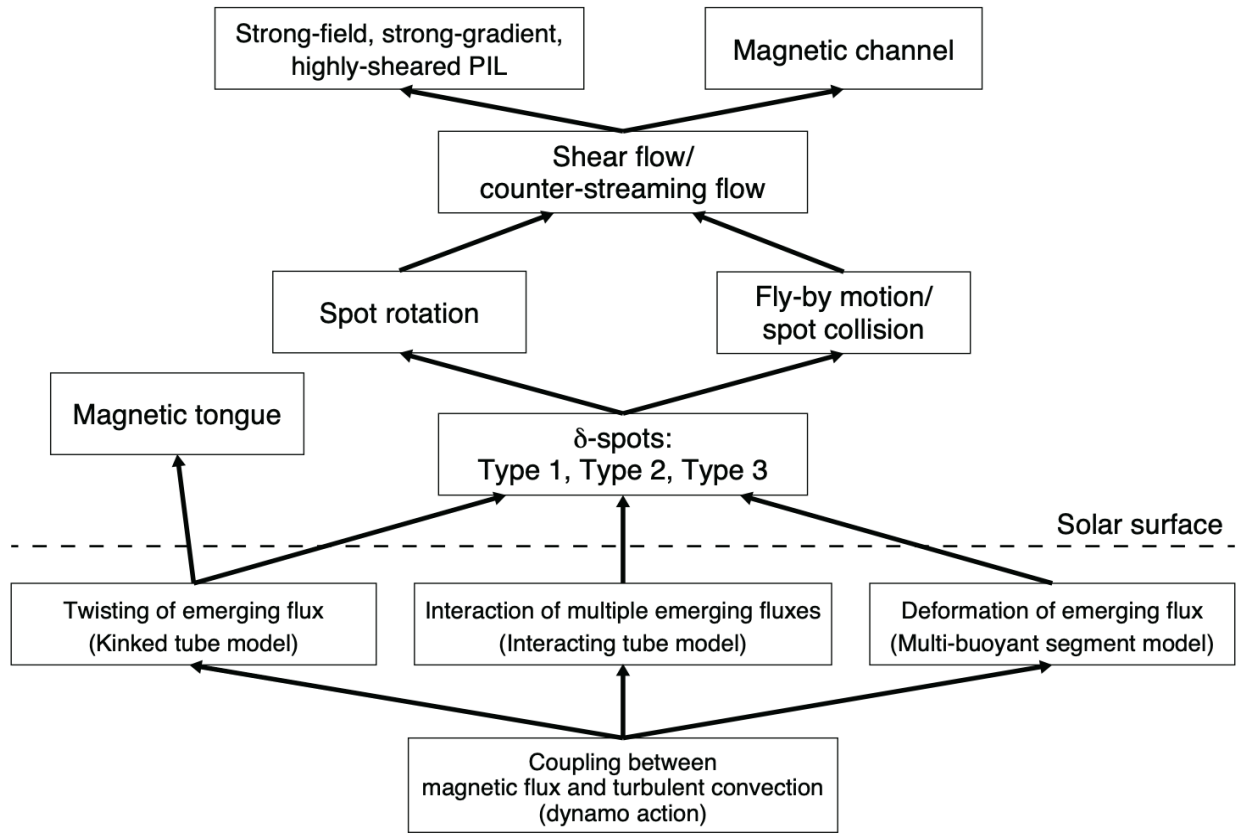}
\caption{Selected key observational features (mainly photospheric) discussed in this article and their relationships as revealed by the series of observational and theoretical studies. The morphological and magnetic complexities of flare-productive ARs are derived ultimately from the close interplay between magnetic flux and turbulent convection, i.e., the magnetoconvective nature of dynamo mechanisms in the deep interior.}
\label{fig:flowchart}
\end{figure*}

In this article, we discussed important observational features of flare-productive ARs, including $\delta$-spots, sheared PILs, magnetic channels, rotating spots, and flux ropes. These structures are reproduced by numerical models that have been performed over the past decades. In particular, state-of-the-art realistic flux emergence simulations that contain large convective cells in deeper layers are now able to reproduce all these features in a self consistent manner, and we understand that the interaction between magnetic flux and turbulent convection introduces such complexity in the ARs (see Fig. \ref{fig:flowchart}).

Although most sophisticated models reveal the critical aspects of the formation of flare-productive ARs, there still remains a scale gap between the localized flux-emergence simulations and global-scale dynamo models. For instance, almost none of the initial flux tubes in the flaring AR models introduced in Sections \ref{sec:sim} and \ref{sec:r2d2} is given as a dynamo-generated toroidal flux, and it is not obvious if these initial conditions are truly natural. Therefore, the next step of the R2D2 calculation would be to test the realistic flux emergence from much deeper layers. Ultimately, it is necessary to simultaneously solve the dynamo action in the deep convection zone of a global scale and the development of ARs on the surface to gain comprehensive understanding of the entire history from the interior to the surface.

From the observational aspect, it has been pointed out that the flare reconnection is triggered by small-scale, perhaps subarcsec-sized, photospheric fields around the PIL \citep[][and references therein]{2017NatAs...1E..85W}. Therefore, to understand the origin and development of flare triggers, unprecedented high-resolution polarimetry as achieved by DKIST \citep{2020SoPh..295..172R} is demanded. Further, UV observations suggest that the charging of flare energy in the atmosphere may be quantified by the non-thermal broadening of coronal emission lines \citep{2009ApJ...691L..99H,2014PASJ...66S..17I}. Such spectroscopic signatures of energy accumulation over the course of AR development would be one of the optimal targets of the next-generation space UV observatories Solar-C (EUVST) \citep{2019SPIE11118E..07S} and MUSE \citep{2020ApJ...888....3D}. Nevertheless, these efforts would be facilitated more by the support of advanced modeling capabilities.

Recent progress in observing stellar flares indicates the occurrence of superflares on solar-type stars, which are 10--1000 times stronger than those on the Sun \citep[e.g.,][]{2012Natur.485..478M}, and this leads to the discussion on the possibility of superflares on the Sun \citep[e.g.,][]{2013A&A...549A..66A,2013PASJ...65...49S}. Then, how do such superflaring starspots occur? As described in Section \ref{sec:obs}, on the Sun, we know that large, complex, rapidly-evolving ARs tend to produce massive flares. Therefore, it is required to obtain the information of starspot evolution and its relation with stellar flares \citep{2019ApJ...871..187N}. Recently, \citet{2020ApJ...902...36T} focused on the Sun-as-a-star spectral irradiance variations of transiting sunspots and revealed that the multi-wavelength monitoring of the stars may help us understand the structures and evolutions of starspots. In future, probably by combining with the numerical models, we may have a more detailed approach on finding the cause of stellar flares.

\section*{Acknowledgments}

We are grateful to the Special Issue Guest Editors (B. Schmieder and C. H. Mandrini) and the anonymous referees for improving the quality of the manuscript.
This work was supported by JSPS KAKENHI Grant Numbers JP18H03723 (PI: Y. Katsukawa), JP20KK0072 (PI: S. Toriumi),  JP21H01124 (PI: T. Yokoyama), and JP21H04492 (PI: K. Kusano), and by the NINS program for cross-disciplinary study (grant Nos. 01321802 and 01311904) on Turbulence, Transport, and Heating Dynamics in Laboratory and Astrophysical Plasmas: ``SoLaBo-X.''

\bibliographystyle{model5-names}
\biboptions{authoryear}
\bibliography{toriumi2021asr}

\begin{thebibliography}{104}
\expandafter\ifx\csname natexlab\endcsname\relax\def\natexlab#1{#1}\fi
\providecommand{\url}[1]{\texttt{#1}}
\providecommand{\href}[2]{#2}
\providecommand{\path}[1]{#1}
\providecommand{\DOIprefix}{doi:}
\providecommand{\ArXivprefix}{arXiv:}
\providecommand{\URLprefix}{URL: }
\providecommand{\Pubmedprefix}{pmid:}
\providecommand{\doi}[1]{\href{http://dx.doi.org/#1}{\path{#1}}}
\providecommand{\Pubmed}[1]{\href{pmid:#1}{\path{#1}}}
\providecommand{\bibinfo}[2]{#2}
\ifx\xfnm\relax \def\xfnm[#1]{\unskip,\space#1}\fi
\bibitem[{{Allen} et~al.(1989){Allen}, {Frank}, {Sauer} \&
  {Reiff}}]{1989EOSTr..70.1479A}
\bibinfo{author}{{Allen}, J.}, \bibinfo{author}{{Frank}, L.},
  \bibinfo{author}{{Sauer}, H.}, \& \bibinfo{author}{{Reiff}, P.}
  (\bibinfo{year}{1989}).
\newblock \bibinfo{title}{{Effects of the March 1989 solar activity}}.
\newblock {\it \bibinfo{journal}{EOS Transactions}\/},  {\it
  \bibinfo{volume}{70}\/}\bibinfo{issue}{(46)}, \bibinfo{pages}{1479--1488}.
  \DOIprefix\doi{10.1029/89EO00409}.
\bibitem[{{Archontis}(2008)}]{2008JGRA..113.3S04A}
\bibinfo{author}{{Archontis}, V.} (\bibinfo{year}{2008}).
\newblock \bibinfo{title}{{Magnetic flux emergence in the Sun}}.
\newblock {\it \bibinfo{journal}{Journal of Geophysical Research (Space
  Physics)}\/},  {\it \bibinfo{volume}{113}\/}\bibinfo{issue}{(A3)},
  \bibinfo{pages}{A03S04}. \DOIprefix\doi{10.1029/2007JA012422}.
\bibitem[{{Aulanier} et~al.(2013){Aulanier}, {D{\'e}moulin}, {Schrijver},
  {Janvier}, {Pariat} \& {Schmieder}}]{2013A&A...549A..66A}
\bibinfo{author}{{Aulanier}, G.}, \bibinfo{author}{{D{\'e}moulin}, P.},
  \bibinfo{author}{{Schrijver}, C.~J.}, \bibinfo{author}{{Janvier}, M.},
  \bibinfo{author}{{Pariat}, E.}, \& \bibinfo{author}{{Schmieder}, B.}
  (\bibinfo{year}{2013}).
\newblock \bibinfo{title}{{The standard flare model in three dimensions. II.
  Upper limit on solar flare energy}}.
\newblock {\it \bibinfo{journal}{Astronomy \& Astrophysics}\/},  {\it
  \bibinfo{volume}{549}\/}, \bibinfo{pages}{A66}.
  \DOIprefix\doi{10.1051/0004-6361/201220406}.
  \href{http://arxiv.org/abs/1212.2086}{\tt arXiv:1212.2086}.
\bibitem[{{Bamba} et~al.(2013){Bamba}, {Kusano}, {Yamamoto} \&
  {Okamoto}}]{2013ApJ...778...48B}
\bibinfo{author}{{Bamba}, Y.}, \bibinfo{author}{{Kusano}, K.},
  \bibinfo{author}{{Yamamoto}, T.~T.}, \& \bibinfo{author}{{Okamoto}, T.~J.}
  (\bibinfo{year}{2013}).
\newblock \bibinfo{title}{{Study on the Triggering Process of Solar Flares
  Based on Hinode/SOT Observations}}.
\newblock {\it \bibinfo{journal}{The Astrophysical Journal}\/},  {\it
  \bibinfo{volume}{778}\/}\bibinfo{issue}{(1)}, \bibinfo{pages}{48}.
  \DOIprefix\doi{10.1088/0004-637X/778/1/48}.
  \href{http://arxiv.org/abs/1309.5465}{\tt arXiv:1309.5465}.
\bibitem[{{Benz}(2017)}]{2017LRSP...14....2B}
\bibinfo{author}{{Benz}, A.~O.} (\bibinfo{year}{2017}).
\newblock \bibinfo{title}{{Flare Observations}}.
\newblock {\it \bibinfo{journal}{Living Reviews in Solar Physics}\/},  {\it
  \bibinfo{volume}{14}\/}\bibinfo{issue}{(1)}, \bibinfo{pages}{2}.
  \DOIprefix\doi{10.1007/s41116-016-0004-3}.
\bibitem[{{Berger} \& {Field}(1984)}]{1984JFM...147..133B}
\bibinfo{author}{{Berger}, M.~A.}, \& \bibinfo{author}{{Field}, G.~B.}
  (\bibinfo{year}{1984}).
\newblock \bibinfo{title}{{The topological properties of magnetic helicity}}.
\newblock {\it \bibinfo{journal}{Journal of Fluid Mechanics}\/},  {\it
  \bibinfo{volume}{147}\/}, \bibinfo{pages}{133--148}.
  \DOIprefix\doi{10.1017/S0022112084002019}.
\bibitem[{{Bobra} \& {Couvidat}(2015)}]{2015ApJ...798..135B}
\bibinfo{author}{{Bobra}, M.~G.}, \& \bibinfo{author}{{Couvidat}, S.}
  (\bibinfo{year}{2015}).
\newblock \bibinfo{title}{{Solar Flare Prediction Using SDO/HMI Vector Magnetic
  Field Data with a Machine-learning Algorithm}}.
\newblock {\it \bibinfo{journal}{The Astrophysical Journal}\/},  {\it
  \bibinfo{volume}{798}\/}\bibinfo{issue}{(2)}, \bibinfo{pages}{135}.
  \DOIprefix\doi{10.1088/0004-637X/798/2/135}.
  \href{http://arxiv.org/abs/1411.1405}{\tt arXiv:1411.1405}.
\bibitem[{{Canfield} et~al.(1999){Canfield}, {Hudson} \&
  {McKenzie}}]{1999GeoRL..26..627C}
\bibinfo{author}{{Canfield}, R.~C.}, \bibinfo{author}{{Hudson}, H.~S.}, \&
  \bibinfo{author}{{McKenzie}, D.~E.} (\bibinfo{year}{1999}).
\newblock \bibinfo{title}{{Sigmoidal morphology and eruptive solar activity}}.
\newblock {\it \bibinfo{journal}{Geophysical Research Letters}\/},  {\it
  \bibinfo{volume}{26}\/}\bibinfo{issue}{(6)}, \bibinfo{pages}{627--630}.
  \DOIprefix\doi{10.1029/1999GL900105}.
\bibitem[{{Carmichael}(1964)}]{1964NASSP..50..451C}
\bibinfo{author}{{Carmichael}, H.} (\bibinfo{year}{1964}).
\newblock \bibinfo{title}{{A Process for Flares}}.
\newblock In {\it \bibinfo{booktitle}{NASA Special Publication}\/} (pp.
  \bibinfo{pages}{451--456}).
\newblock volume~\bibinfo{volume}{50}.
\bibitem[{{Carrington}(1859)}]{1859MNRAS..20...13C}
\bibinfo{author}{{Carrington}, R.~C.} (\bibinfo{year}{1859}).
\newblock \bibinfo{title}{{Description of a Singular Appearance seen in the Sun
  on September 1, 1859}}.
\newblock {\it \bibinfo{journal}{Monthly Notices of the Royal Astronomical
  Society}\/},  {\it \bibinfo{volume}{20}\/}, \bibinfo{pages}{13--15}.
  \DOIprefix\doi{10.1093/mnras/20.1.13}.
\bibitem[{{Castellanos Dur{\'a}n} et~al.(2020){Castellanos Dur{\'a}n}, {Lagg},
  {Solanki} \& {van Noort}}]{2020ApJ...895..129C}
\bibinfo{author}{{Castellanos Dur{\'a}n}, J.~S.}, \bibinfo{author}{{Lagg}, A.},
  \bibinfo{author}{{Solanki}, S.~K.}, \& \bibinfo{author}{{van Noort}, M.}
  (\bibinfo{year}{2020}).
\newblock \bibinfo{title}{{Detection of the Strongest Magnetic Field in a
  Sunspot Light Bridge}}.
\newblock {\it \bibinfo{journal}{The Astrophysical Journal}\/},  {\it
  \bibinfo{volume}{895}\/}\bibinfo{issue}{(2)}, \bibinfo{pages}{129}.
  \DOIprefix\doi{10.3847/1538-4357/ab83f1}.
  \href{http://arxiv.org/abs/2003.12078}{\tt arXiv:2003.12078}.
\bibitem[{{Chen} et~al.(2017){Chen}, {Rempel} \& {Fan}}]{2017ApJ...846..149C}
\bibinfo{author}{{Chen}, F.}, \bibinfo{author}{{Rempel}, M.}, \&
  \bibinfo{author}{{Fan}, Y.} (\bibinfo{year}{2017}).
\newblock \bibinfo{title}{{Emergence of Magnetic Flux Generated in a Solar
  Convective Dynamo. I. The Formation of Sunspots and Active Regions, and The
  Origin of Their Asymmetries}}.
\newblock {\it \bibinfo{journal}{The Astrophysical Journal}\/},  {\it
  \bibinfo{volume}{846}\/}\bibinfo{issue}{(2)}, \bibinfo{pages}{149}.
  \DOIprefix\doi{10.3847/1538-4357/aa85a0}.
  \href{http://arxiv.org/abs/1704.05999}{\tt arXiv:1704.05999}.
\bibitem[{{Chen}(2011)}]{2011LRSP....8....1C}
\bibinfo{author}{{Chen}, P.~F.} (\bibinfo{year}{2011}).
\newblock \bibinfo{title}{{Coronal Mass Ejections: Models and Their
  Observational Basis}}.
\newblock {\it \bibinfo{journal}{Living Reviews in Solar Physics}\/},  {\it
  \bibinfo{volume}{8}\/}\bibinfo{issue}{(1)}, \bibinfo{pages}{1}.
  \DOIprefix\doi{10.12942/lrsp-2011-1}.
\bibitem[{{Cheung} \& {Isobe}(2014)}]{2014LRSP...11....3C}
\bibinfo{author}{{Cheung}, M. C.~M.}, \& \bibinfo{author}{{Isobe}, H.}
  (\bibinfo{year}{2014}).
\newblock \bibinfo{title}{{Flux Emergence (Theory)}}.
\newblock {\it \bibinfo{journal}{Living Reviews in Solar Physics}\/},  {\it
  \bibinfo{volume}{11}\/}\bibinfo{issue}{(1)}, \bibinfo{pages}{3}.
  \DOIprefix\doi{10.12942/lrsp-2014-3}.
\bibitem[{{Cheung} et~al.(2019){Cheung}, {Rempel}, {Chintzoglou}, {Chen},
  {Testa}, {Mart{\'\i}nez-Sykora}, {Sainz Dalda}, {DeRosa}, {Malanushenko},
  {Hansteen}, {De Pontieu}, {Carlsson}, {Gudiksen} \&
  {McIntosh}}]{2019NatAs...3..160C}
\bibinfo{author}{{Cheung}, M.~C.~M.}, \bibinfo{author}{{Rempel}, M.},
  \bibinfo{author}{{Chintzoglou}, G.}, \bibinfo{author}{{Chen}, F.},
  \bibinfo{author}{{Testa}, P.}, \bibinfo{author}{{Mart{\'\i}nez-Sykora}, J.},
  \bibinfo{author}{{Sainz Dalda}, A.}, \bibinfo{author}{{DeRosa}, M.~L.},
  \bibinfo{author}{{Malanushenko}, A.}, \bibinfo{author}{{Hansteen}, V.},
  \bibinfo{author}{{De Pontieu}, B.}, \bibinfo{author}{{Carlsson}, M.},
  \bibinfo{author}{{Gudiksen}, B.}, \& \bibinfo{author}{{McIntosh}, S.~W.}
  (\bibinfo{year}{2019}).
\newblock \bibinfo{title}{{A comprehensive three-dimensional radiative
  magnetohydrodynamic simulation of a solar flare}}.
\newblock {\it \bibinfo{journal}{Nature Astronomy}\/},  {\it
  \bibinfo{volume}{3}\/}, \bibinfo{pages}{160--166}.
  \DOIprefix\doi{10.1038/s41550-018-0629-3}.
\bibitem[{{Cheung} et~al.(2010){Cheung}, {Rempel}, {Title} \&
  {Sch{\"u}ssler}}]{2010ApJ...720..233C}
\bibinfo{author}{{Cheung}, M.~C.~M.}, \bibinfo{author}{{Rempel}, M.},
  \bibinfo{author}{{Title}, A.~M.}, \& \bibinfo{author}{{Sch{\"u}ssler}, M.}
  (\bibinfo{year}{2010}).
\newblock \bibinfo{title}{{Simulation of the Formation of a Solar Active
  Region}}.
\newblock {\it \bibinfo{journal}{The Astrophysical Journal}\/},  {\it
  \bibinfo{volume}{720}\/}\bibinfo{issue}{(1)}, \bibinfo{pages}{233--244}.
  \DOIprefix\doi{10.1088/0004-637X/720/1/233}.
  \href{http://arxiv.org/abs/1006.4117}{\tt arXiv:1006.4117}.
\bibitem[{{Cheung} et~al.(2008){Cheung}, {Sch{\"u}ssler}, {Tarbell} \&
  {Title}}]{2008ApJ...687.1373C}
\bibinfo{author}{{Cheung}, M.~C.~M.}, \bibinfo{author}{{Sch{\"u}ssler}, M.},
  \bibinfo{author}{{Tarbell}, T.~D.}, \& \bibinfo{author}{{Title}, A.~M.}
  (\bibinfo{year}{2008}).
\newblock \bibinfo{title}{{Solar Surface Emerging Flux Regions: A Comparative
  Study of Radiative MHD Modeling and Hinode SOT Observations}}.
\newblock {\it \bibinfo{journal}{The Astrophysical Journal}\/},  {\it
  \bibinfo{volume}{687}\/}\bibinfo{issue}{(2)}, \bibinfo{pages}{1373--1387}.
  \DOIprefix\doi{10.1086/591245}. \href{http://arxiv.org/abs/0810.5723}{\tt
  arXiv:0810.5723}.
\bibitem[{{Chintzoglou} \& {Zhang}(2013)}]{2013ApJ...764L...3C}
\bibinfo{author}{{Chintzoglou}, G.}, \& \bibinfo{author}{{Zhang}, J.}
  (\bibinfo{year}{2013}).
\newblock \bibinfo{title}{{Reconstructing the Subsurface Three-dimensional
  Magnetic Structure of a Solar Active Region Using SDO/HMI Observations}}.
\newblock {\it \bibinfo{journal}{The Astrophysical Journal Letters}\/},  {\it
  \bibinfo{volume}{764}\/}\bibinfo{issue}{(1)}, \bibinfo{pages}{L3}.
  \DOIprefix\doi{10.1088/2041-8205/764/1/L3}.
  \href{http://arxiv.org/abs/1301.4651}{\tt arXiv:1301.4651}.
\bibitem[{{Chintzoglou} et~al.(2019){Chintzoglou}, {Zhang}, {Cheung} \&
  {Kazachenko}}]{2019ApJ...871...67C}
\bibinfo{author}{{Chintzoglou}, G.}, \bibinfo{author}{{Zhang}, J.},
  \bibinfo{author}{{Cheung}, M. C.~M.}, \& \bibinfo{author}{{Kazachenko}, M.}
  (\bibinfo{year}{2019}).
\newblock \bibinfo{title}{{The Origin of Major Solar Activity: Collisional
  Shearing between Nonconjugated Polarities of Multiple Bipoles Emerging within
  Active Regions}}.
\newblock {\it \bibinfo{journal}{The Astrophysical Journal}\/},  {\it
  \bibinfo{volume}{871}\/}\bibinfo{issue}{(1)}, \bibinfo{pages}{67}.
  \DOIprefix\doi{10.3847/1538-4357/aaef30}.
  \href{http://arxiv.org/abs/1811.02186}{\tt arXiv:1811.02186}.
\bibitem[{{De Pontieu} et~al.(2020){De Pontieu}, {Mart{\'\i}nez-Sykora},
  {Testa}, {Winebarger}, {Daw}, {Hansteen}, {Cheung} \&
  {Antolin}}]{2020ApJ...888....3D}
\bibinfo{author}{{De Pontieu}, B.}, \bibinfo{author}{{Mart{\'\i}nez-Sykora},
  J.}, \bibinfo{author}{{Testa}, P.}, \bibinfo{author}{{Winebarger}, A.~R.},
  \bibinfo{author}{{Daw}, A.}, \bibinfo{author}{{Hansteen}, V.},
  \bibinfo{author}{{Cheung}, M. C.~M.}, \& \bibinfo{author}{{Antolin}, P.}
  (\bibinfo{year}{2020}).
\newblock \bibinfo{title}{{The Multi-slit Approach to Coronal Spectroscopy with
  the Multi-slit Solar Explorer (MUSE)}}.
\newblock {\it \bibinfo{journal}{The Astrophysical Journal}\/},  {\it
  \bibinfo{volume}{888}\/}\bibinfo{issue}{(1)}, \bibinfo{pages}{3}.
  \DOIprefix\doi{10.3847/1538-4357/ab5b03}.
  \href{http://arxiv.org/abs/1909.08818}{\tt arXiv:1909.08818}.
\bibitem[{{D{\'e}moulin} \& {Pariat}(2009)}]{2009AdSpR..43.1013D}
\bibinfo{author}{{D{\'e}moulin}, P.}, \& \bibinfo{author}{{Pariat}, E.}
  (\bibinfo{year}{2009}).
\newblock \bibinfo{title}{{Modelling and observations of photospheric magnetic
  helicity}}.
\newblock {\it \bibinfo{journal}{Advances in Space Research}\/},  {\it
  \bibinfo{volume}{43}\/}\bibinfo{issue}{(7)}, \bibinfo{pages}{1013--1031}.
  \DOIprefix\doi{10.1016/j.asr.2008.12.004}.
\bibitem[{{Elsasser}(1956)}]{1956RvMP...28..135E}
\bibinfo{author}{{Elsasser}, W.~M.} (\bibinfo{year}{1956}).
\newblock \bibinfo{title}{{Hydromagnetic Dynamo Theory}}.
\newblock {\it \bibinfo{journal}{Reviews of Modern Physics}\/},  {\it
  \bibinfo{volume}{28}\/}\bibinfo{issue}{(2)}, \bibinfo{pages}{135--163}.
  \DOIprefix\doi{10.1103/RevModPhys.28.135}.
\bibitem[{{Falconer} et~al.(2002){Falconer}, {Moore} \&
  {Gary}}]{2002ApJ...569.1016F}
\bibinfo{author}{{Falconer}, D.~A.}, \bibinfo{author}{{Moore}, R.~L.}, \&
  \bibinfo{author}{{Gary}, G.~A.} (\bibinfo{year}{2002}).
\newblock \bibinfo{title}{{Correlation of the Coronal Mass Ejection
  Productivity of Solar Active Regions with Measures of Their Global
  Nonpotentiality from Vector Magnetograms: Baseline Results}}.
\newblock {\it \bibinfo{journal}{The Astrophysical Journal}\/},  {\it
  \bibinfo{volume}{569}\/}\bibinfo{issue}{(2)}, \bibinfo{pages}{1016--1025}.
  \DOIprefix\doi{10.1086/339161}.
\bibitem[{{Fan}(2009)}]{2009LRSP....6....4F}
\bibinfo{author}{{Fan}, Y.} (\bibinfo{year}{2009}).
\newblock \bibinfo{title}{{Magnetic Fields in the Solar Convection Zone}}.
\newblock {\it \bibinfo{journal}{Living Reviews in Solar Physics}\/},  {\it
  \bibinfo{volume}{6}\/}\bibinfo{issue}{(1)}, \bibinfo{pages}{4}.
  \DOIprefix\doi{10.12942/lrsp-2009-4}.
\bibitem[{{Fan} et~al.(1999){Fan}, {Zweibel}, {Linton} \&
  {Fisher}}]{1999ApJ...521..460F}
\bibinfo{author}{{Fan}, Y.}, \bibinfo{author}{{Zweibel}, E.~G.},
  \bibinfo{author}{{Linton}, M.~G.}, \& \bibinfo{author}{{Fisher}, G.~H.}
  (\bibinfo{year}{1999}).
\newblock \bibinfo{title}{{The Rise of Kink-unstable Magnetic Flux Tubes and
  the Origin of {\ensuremath{\delta}}-Configuration Sunspots}}.
\newblock {\it \bibinfo{journal}{The Astrophysical Journal}\/},  {\it
  \bibinfo{volume}{521}\/}\bibinfo{issue}{(1)}, \bibinfo{pages}{460--477}.
  \DOIprefix\doi{10.1086/307533}.
\bibitem[{{Fang} \& {Fan}(2015)}]{2015ApJ...806...79F}
\bibinfo{author}{{Fang}, F.}, \& \bibinfo{author}{{Fan}, Y.}
  (\bibinfo{year}{2015}).
\newblock \bibinfo{title}{{{\ensuremath{\delta}}-Sunspot Formation in
  Simulation of Active-region-scale Flux Emergence}}.
\newblock {\it \bibinfo{journal}{The Astrophysical Journal}\/},  {\it
  \bibinfo{volume}{806}\/}\bibinfo{issue}{(1)}, \bibinfo{pages}{79}.
  \DOIprefix\doi{10.1088/0004-637X/806/1/79}.
  \href{http://arxiv.org/abs/1504.04393}{\tt arXiv:1504.04393}.
\bibitem[{{Finn} \& {Antonsen Jr}(1985)}]{finn1985}
\bibinfo{author}{{Finn}, J.~M.}, \& \bibinfo{author}{{Antonsen Jr}, T.~M.}
  (\bibinfo{year}{1985}).
\newblock \bibinfo{title}{{Magnetic helicity: What is it, and what is it good
  for?}}
\newblock {\it \bibinfo{journal}{Comments on Plasma Physics and Controlled
  Fusion}\/},  {\it \bibinfo{volume}{9}\/}, \bibinfo{pages}{111--126}.
\bibitem[{{Fletcher} et~al.(2011){Fletcher}, {Dennis}, {Hudson}, {Krucker},
  {Phillips}, {Veronig}, {Battaglia}, {Bone}, {Caspi}, {Chen}, {Gallagher},
  {Grigis}, {Ji}, {Liu}, {Milligan} \& {Temmer}}]{2011SSRv..159...19F}
\bibinfo{author}{{Fletcher}, L.}, \bibinfo{author}{{Dennis}, B.~R.},
  \bibinfo{author}{{Hudson}, H.~S.}, \bibinfo{author}{{Krucker}, S.},
  \bibinfo{author}{{Phillips}, K.}, \bibinfo{author}{{Veronig}, A.},
  \bibinfo{author}{{Battaglia}, M.}, \bibinfo{author}{{Bone}, L.},
  \bibinfo{author}{{Caspi}, A.}, \bibinfo{author}{{Chen}, Q.},
  \bibinfo{author}{{Gallagher}, P.}, \bibinfo{author}{{Grigis}, P.~T.},
  \bibinfo{author}{{Ji}, H.}, \bibinfo{author}{{Liu}, W.},
  \bibinfo{author}{{Milligan}, R.~O.}, \& \bibinfo{author}{{Temmer}, M.}
  (\bibinfo{year}{2011}).
\newblock \bibinfo{title}{{An Observational Overview of Solar Flares}}.
\newblock {\it \bibinfo{journal}{Space Science Reviews}\/},  {\it
  \bibinfo{volume}{159}\/}\bibinfo{issue}{(1-4)}, \bibinfo{pages}{19--106}.
  \DOIprefix\doi{10.1007/s11214-010-9701-8}.
  \href{http://arxiv.org/abs/1109.5932}{\tt arXiv:1109.5932}.
\bibitem[{{Hagyard} et~al.(1984){Hagyard}, {Smith}, {Teuber} \&
  {West}}]{1984SoPh...91..115H}
\bibinfo{author}{{Hagyard}, M.~J.}, \bibinfo{author}{{Smith}, J., J.~B.},
  \bibinfo{author}{{Teuber}, D.}, \& \bibinfo{author}{{West}, E.~A.}
  (\bibinfo{year}{1984}).
\newblock \bibinfo{title}{{A Quantitative Study Relating Observed Shear in
  Photospheric Magnetic Fields to Repeated Flaring}}.
\newblock {\it \bibinfo{journal}{Solar Physics}\/},  {\it
  \bibinfo{volume}{91}\/}\bibinfo{issue}{(1)}, \bibinfo{pages}{115--126}.
  \DOIprefix\doi{10.1007/BF00213618}.
\bibitem[{{Harra} et~al.(2009){Harra}, {Williams}, {Wallace}, {Magara}, {Hara},
  {Tsuneta}, {Sterling} \& {Doschek}}]{2009ApJ...691L..99H}
\bibinfo{author}{{Harra}, L.~K.}, \bibinfo{author}{{Williams}, D.~R.},
  \bibinfo{author}{{Wallace}, A.~J.}, \bibinfo{author}{{Magara}, T.},
  \bibinfo{author}{{Hara}, H.}, \bibinfo{author}{{Tsuneta}, S.},
  \bibinfo{author}{{Sterling}, A.~C.}, \& \bibinfo{author}{{Doschek}, G.~A.}
  (\bibinfo{year}{2009}).
\newblock \bibinfo{title}{{Coronal Nonthermal Velocity Following Helicity
  Injection Before an X-Class Flare}}.
\newblock {\it \bibinfo{journal}{The Astrophysical Journal Letters}\/},  {\it
  \bibinfo{volume}{691}\/}\bibinfo{issue}{(2)}, \bibinfo{pages}{L99--L102}.
  \DOIprefix\doi{10.1088/0004-637X/691/2/L99}.
\bibitem[{{Harvey} \& {Harvey}(1976)}]{1976SoPh...47..233H}
\bibinfo{author}{{Harvey}, K.~L.}, \& \bibinfo{author}{{Harvey}, J.~W.}
  (\bibinfo{year}{1976}).
\newblock \bibinfo{title}{{A study of the magnetic and velocity fields in an
  active region.}}
\newblock {\it \bibinfo{journal}{Solar Physics}\/},  {\it
  \bibinfo{volume}{47}\/}\bibinfo{issue}{(1)}, \bibinfo{pages}{233--246}.
  \DOIprefix\doi{10.1007/BF00152261}.
\bibitem[{{Hinode Review Team} et~al.(2019){Hinode Review Team}, {Al-Janabi},
  {Antolin}, {Baker}, {Bellot Rubio}, {Bradley}, {Brooks}, {Centeno},
  {Culhane}, {Del Zanna}, {Doschek}, {Fletcher}, {Hara}, {Harra}, {Hillier},
  {Imada}, {Klimchuk}, {Mariska}, {Pereira}, {Reeves}, {Sakao}, {Sakurai},
  {Shimizu}, {Shimojo}, {Shiota}, {Solanki}, {Sterling}, {Su}, {Suematsu},
  {Tarbell}, {Tiwari}, {Toriumi}, {Ugarte-Urra}, {Warren}, {Watanabe} \&
  {Young}}]{2019PASJ...71R...1H}
\bibinfo{author}{{Hinode Review Team}}, \bibinfo{author}{{Al-Janabi}, K.},
  \bibinfo{author}{{Antolin}, P.}, \bibinfo{author}{{Baker}, D.},
  \bibinfo{author}{{Bellot Rubio}, L.~R.}, \bibinfo{author}{{Bradley}, L.},
  \bibinfo{author}{{Brooks}, D.~H.}, \bibinfo{author}{{Centeno}, R.},
  \bibinfo{author}{{Culhane}, J.~L.}, \bibinfo{author}{{Del Zanna}, G.},
  \bibinfo{author}{{Doschek}, G.~A.}, \bibinfo{author}{{Fletcher}, L.},
  \bibinfo{author}{{Hara}, H.}, \bibinfo{author}{{Harra}, L.~K.},
  \bibinfo{author}{{Hillier}, A.~S.}, \bibinfo{author}{{Imada}, S.},
  \bibinfo{author}{{Klimchuk}, J.~A.}, \bibinfo{author}{{Mariska}, J.~T.},
  \bibinfo{author}{{Pereira}, T. M.~D.}, \bibinfo{author}{{Reeves}, K.~K.},
  \bibinfo{author}{{Sakao}, T.}, \bibinfo{author}{{Sakurai}, T.},
  \bibinfo{author}{{Shimizu}, T.}, \bibinfo{author}{{Shimojo}, M.},
  \bibinfo{author}{{Shiota}, D.}, \bibinfo{author}{{Solanki}, S.~K.},
  \bibinfo{author}{{Sterling}, A.~C.}, \bibinfo{author}{{Su}, Y.},
  \bibinfo{author}{{Suematsu}, Y.}, \bibinfo{author}{{Tarbell}, T.~D.},
  \bibinfo{author}{{Tiwari}, S.~K.}, \bibinfo{author}{{Toriumi}, S.},
  \bibinfo{author}{{Ugarte-Urra}, I.}, \bibinfo{author}{{Warren}, H.~P.},
  \bibinfo{author}{{Watanabe}, T.}, \& \bibinfo{author}{{Young}, P.~R.}
  (\bibinfo{year}{2019}).
\newblock \bibinfo{title}{{Achievements of Hinode in the first eleven years}}.
\newblock {\it \bibinfo{journal}{Publications of the Astronomical Society of
  Japan}\/},  {\it \bibinfo{volume}{71}\/}\bibinfo{issue}{(5)},
  \bibinfo{pages}{R1}. \DOIprefix\doi{10.1093/pasj/psz084}.
\bibitem[{{Hirayama}(1974)}]{1974SoPh...34..323H}
\bibinfo{author}{{Hirayama}, T.} (\bibinfo{year}{1974}).
\newblock \bibinfo{title}{{Theoretical Model of Flares and Prominences. I:
  Evaporating Flare Model}}.
\newblock {\it \bibinfo{journal}{Solar Physics}\/},  {\it
  \bibinfo{volume}{34}\/}\bibinfo{issue}{(2)}, \bibinfo{pages}{323--338}.
  \DOIprefix\doi{10.1007/BF00153671}.
\bibitem[{{Hodgson}(1859)}]{1859MNRAS..20...15H}
\bibinfo{author}{{Hodgson}, R.} (\bibinfo{year}{1859}).
\newblock \bibinfo{title}{{On a curious Appearance seen in the Sun}}.
\newblock {\it \bibinfo{journal}{Monthly Notices of the Royal Astronomical
  Society}\/},  {\it \bibinfo{volume}{20}\/}, \bibinfo{pages}{15--16}.
  \DOIprefix\doi{10.1093/mnras/20.1.15}.
\bibitem[{{Hotta} \& {Iijima}(2020)}]{2020MNRAS.494.2523H}
\bibinfo{author}{{Hotta}, H.}, \& \bibinfo{author}{{Iijima}, H.}
  (\bibinfo{year}{2020}).
\newblock \bibinfo{title}{{On rising magnetic flux tube and formation of
  sunspots in a deep domain}}.
\newblock {\it \bibinfo{journal}{Monthly Notices of the Royal Astronomical
  Society}\/},  {\it \bibinfo{volume}{494}\/}\bibinfo{issue}{(2)},
  \bibinfo{pages}{2523--2537}. \DOIprefix\doi{10.1093/mnras/staa844}.
  \href{http://arxiv.org/abs/2003.10583}{\tt arXiv:2003.10583}.
\bibitem[{{Hotta} et~al.(2019){Hotta}, {Iijima} \&
  {Kusano}}]{2019SciA....5.2307H}
\bibinfo{author}{{Hotta}, H.}, \bibinfo{author}{{Iijima}, H.}, \&
  \bibinfo{author}{{Kusano}, K.} (\bibinfo{year}{2019}).
\newblock \bibinfo{title}{{Weak influence of near-surface layer on solar deep
  convection zone revealed by comprehensive simulation from base to surface}}.
\newblock {\it \bibinfo{journal}{Science Advances}\/},  {\it
  \bibinfo{volume}{5}\/}\bibinfo{issue}{(1)}, \bibinfo{pages}{2307}.
  \DOIprefix\doi{10.1126/sciadv.aau2307}.
\bibitem[{{Hotta} et~al.(2012){Hotta}, {Rempel}, {Yokoyama}, {Iida} \&
  {Fan}}]{2012A&A...539A..30H}
\bibinfo{author}{{Hotta}, H.}, \bibinfo{author}{{Rempel}, M.},
  \bibinfo{author}{{Yokoyama}, T.}, \bibinfo{author}{{Iida}, Y.}, \&
  \bibinfo{author}{{Fan}, Y.} (\bibinfo{year}{2012}).
\newblock \bibinfo{title}{{Numerical calculation of convection with reduced
  speed of sound technique}}.
\newblock {\it \bibinfo{journal}{Astronomy \& Astrophysics}\/},  {\it
  \bibinfo{volume}{539}\/}, \bibinfo{pages}{A30}.
  \DOIprefix\doi{10.1051/0004-6361/201118268}.
  \href{http://arxiv.org/abs/1201.1061}{\tt arXiv:1201.1061}.
\bibitem[{{Hotta} \& {Toriumi}(2020)}]{2020MNRAS.498.2925H}
\bibinfo{author}{{Hotta}, H.}, \& \bibinfo{author}{{Toriumi}, S.}
  (\bibinfo{year}{2020}).
\newblock \bibinfo{title}{{Formation of superstrong horizontal magnetic field
  in delta-type sunspot in radiation magnetohydrodynamic simulations}}.
\newblock {\it \bibinfo{journal}{Monthly Notices of the Royal Astronomical
  Society}\/},  {\it \bibinfo{volume}{498}\/}\bibinfo{issue}{(2)},
  \bibinfo{pages}{2925--2935}. \DOIprefix\doi{10.1093/mnras/staa2529}.
  \href{http://arxiv.org/abs/2008.07741}{\tt arXiv:2008.07741}.
\bibitem[{{Imada} et~al.(2014){Imada}, {Bamba} \&
  {Kusano}}]{2014PASJ...66S..17I}
\bibinfo{author}{{Imada}, S.}, \bibinfo{author}{{Bamba}, Y.}, \&
  \bibinfo{author}{{Kusano}, K.} (\bibinfo{year}{2014}).
\newblock \bibinfo{title}{{Coronal behavior before the large flare onset}}.
\newblock {\it \bibinfo{journal}{Publications of the Astronomical Society of
  Japan}\/},  {\it \bibinfo{volume}{66}\/}, \bibinfo{pages}{S17}.
  \DOIprefix\doi{10.1093/pasj/psu092}.
  \href{http://arxiv.org/abs/1408.2585}{\tt arXiv:1408.2585}.
\bibitem[{{Inoue} et~al.(2018){Inoue}, {Kusano}, {B{\"u}chner} \&
  {Sk{\'a}la}}]{2018NatCo...9..174I}
\bibinfo{author}{{Inoue}, S.}, \bibinfo{author}{{Kusano}, K.},
  \bibinfo{author}{{B{\"u}chner}, J.}, \& \bibinfo{author}{{Sk{\'a}la}, J.}
  (\bibinfo{year}{2018}).
\newblock \bibinfo{title}{{Formation and dynamics of a solar eruptive flux
  tube}}.
\newblock {\it \bibinfo{journal}{Nature Communications}\/},  {\it
  \bibinfo{volume}{9}\/}, \bibinfo{pages}{174}.
  \DOIprefix\doi{10.1038/s41467-017-02616-8}.
\bibitem[{{Janvier} et~al.(2015){Janvier}, {Aulanier} \&
  {D{\'e}moulin}}]{2015SoPh..290.3425J}
\bibinfo{author}{{Janvier}, M.}, \bibinfo{author}{{Aulanier}, G.}, \&
  \bibinfo{author}{{D{\'e}moulin}, P.} (\bibinfo{year}{2015}).
\newblock \bibinfo{title}{{From Coronal Observations to MHD Simulations, the
  Building Blocks for 3D Models of Solar Flares (Invited Review)}}.
\newblock {\it \bibinfo{journal}{Solar Physics}\/},  {\it
  \bibinfo{volume}{290}\/}\bibinfo{issue}{(12)}, \bibinfo{pages}{3425--3456}.
  \DOIprefix\doi{10.1007/s11207-015-0710-3}.
  \href{http://arxiv.org/abs/1505.05299}{\tt arXiv:1505.05299}.
\bibitem[{{Jouve} et~al.(2013){Jouve}, {Brun} \&
  {Aulanier}}]{2013ApJ...762....4J}
\bibinfo{author}{{Jouve}, L.}, \bibinfo{author}{{Brun}, A.~S.}, \&
  \bibinfo{author}{{Aulanier}, G.} (\bibinfo{year}{2013}).
\newblock \bibinfo{title}{{Global dynamics of subsurface solar active
  regions}}.
\newblock {\it \bibinfo{journal}{The Astrophysical Journal}\/},  {\it
  \bibinfo{volume}{762}\/}\bibinfo{issue}{(1)}, \bibinfo{pages}{4}.
  \DOIprefix\doi{10.1088/0004-637X/762/1/4}.
  \href{http://arxiv.org/abs/1211.7251}{\tt arXiv:1211.7251}.
\bibitem[{{Jouve} et~al.(2018){Jouve}, {Brun} \&
  {Aulanier}}]{2018ApJ...857...83J}
\bibinfo{author}{{Jouve}, L.}, \bibinfo{author}{{Brun}, A.~S.}, \&
  \bibinfo{author}{{Aulanier}, G.} (\bibinfo{year}{2018}).
\newblock \bibinfo{title}{{Interactions of Twisted {\ensuremath{\Omega}}-loops
  in a Model Solar Convection Zone}}.
\newblock {\it \bibinfo{journal}{The Astrophysical Journal}\/},  {\it
  \bibinfo{volume}{857}\/}\bibinfo{issue}{(2)}, \bibinfo{pages}{83}.
  \DOIprefix\doi{10.3847/1538-4357/aab5b6}.
  \href{http://arxiv.org/abs/1803.04709}{\tt arXiv:1803.04709}.
\bibitem[{{Knizhnik} et~al.(2018){Knizhnik}, {Linton} \&
  {DeVore}}]{2018ApJ...864...89K}
\bibinfo{author}{{Knizhnik}, K.~J.}, \bibinfo{author}{{Linton}, M.~G.}, \&
  \bibinfo{author}{{DeVore}, C.~R.} (\bibinfo{year}{2018}).
\newblock \bibinfo{title}{{The Role of Twist in Kinked Flux Rope Emergence and
  Delta-spot Formation}}.
\newblock {\it \bibinfo{journal}{The Astrophysical Journal}\/},  {\it
  \bibinfo{volume}{864}\/}\bibinfo{issue}{(1)}, \bibinfo{pages}{89}.
  \DOIprefix\doi{10.3847/1538-4357/aad68c}.
  \href{http://arxiv.org/abs/1808.05562}{\tt arXiv:1808.05562}.
\bibitem[{{Kopp} \& {Pneuman}(1976)}]{1976SoPh...50...85K}
\bibinfo{author}{{Kopp}, R.~A.}, \& \bibinfo{author}{{Pneuman}, G.~W.}
  (\bibinfo{year}{1976}).
\newblock \bibinfo{title}{{Magnetic reconnection in the corona and the loop
  prominence phenomenon.}}
\newblock {\it \bibinfo{journal}{Solar Physics}\/},  {\it
  \bibinfo{volume}{50}\/}\bibinfo{issue}{(1)}, \bibinfo{pages}{85--98}.
  \DOIprefix\doi{10.1007/BF00206193}.
\bibitem[{{Kosugi} et~al.(2007){Kosugi}, {Matsuzaki}, {Sakao}, {Shimizu},
  {Sone}, {Tachikawa}, {Hashimoto}, {Minesugi}, {Ohnishi}, {Yamada}, {Tsuneta},
  {Hara}, {Ichimoto}, {Suematsu}, {Shimojo}, {Watanabe}, {Shimada}, {Davis},
  {Hill}, {Owens}, {Title}, {Culhane}, {Harra}, {Doschek} \&
  {Golub}}]{2007SoPh..243....3K}
\bibinfo{author}{{Kosugi}, T.}, \bibinfo{author}{{Matsuzaki}, K.},
  \bibinfo{author}{{Sakao}, T.}, \bibinfo{author}{{Shimizu}, T.},
  \bibinfo{author}{{Sone}, Y.}, \bibinfo{author}{{Tachikawa}, S.},
  \bibinfo{author}{{Hashimoto}, T.}, \bibinfo{author}{{Minesugi}, K.},
  \bibinfo{author}{{Ohnishi}, A.}, \bibinfo{author}{{Yamada}, T.},
  \bibinfo{author}{{Tsuneta}, S.}, \bibinfo{author}{{Hara}, H.},
  \bibinfo{author}{{Ichimoto}, K.}, \bibinfo{author}{{Suematsu}, Y.},
  \bibinfo{author}{{Shimojo}, M.}, \bibinfo{author}{{Watanabe}, T.},
  \bibinfo{author}{{Shimada}, S.}, \bibinfo{author}{{Davis}, J.~M.},
  \bibinfo{author}{{Hill}, L.~D.}, \bibinfo{author}{{Owens}, J.~K.},
  \bibinfo{author}{{Title}, A.~M.}, \bibinfo{author}{{Culhane}, J.~L.},
  \bibinfo{author}{{Harra}, L.~K.}, \bibinfo{author}{{Doschek}, G.~A.}, \&
  \bibinfo{author}{{Golub}, L.} (\bibinfo{year}{2007}).
\newblock \bibinfo{title}{{The Hinode (Solar-B) Mission: An Overview}}.
\newblock {\it \bibinfo{journal}{Solar Physics}\/},  {\it
  \bibinfo{volume}{243}\/}\bibinfo{issue}{(1)}, \bibinfo{pages}{3--17}.
  \DOIprefix\doi{10.1007/s11207-007-9014-6}.
\bibitem[{{Kubo} et~al.(2007){Kubo}, {Yokoyama}, {Katsukawa}, {Lites},
  {Tsuneta}, {Suematsu}, {Ichimoto}, {Shimizu}, {Nagata}, {Tarbell}, {Shine},
  {Title} \& {Elmore David}}]{2007PASJ...59S.779K}
\bibinfo{author}{{Kubo}, M.}, \bibinfo{author}{{Yokoyama}, T.},
  \bibinfo{author}{{Katsukawa}, Y.}, \bibinfo{author}{{Lites}, B.},
  \bibinfo{author}{{Tsuneta}, S.}, \bibinfo{author}{{Suematsu}, Y.},
  \bibinfo{author}{{Ichimoto}, K.}, \bibinfo{author}{{Shimizu}, T.},
  \bibinfo{author}{{Nagata}, S.}, \bibinfo{author}{{Tarbell}, T.~D.},
  \bibinfo{author}{{Shine}, R.~A.}, \bibinfo{author}{{Title}, A.~M.}, \&
  \bibinfo{author}{{Elmore David}} (\bibinfo{year}{2007}).
\newblock \bibinfo{title}{{Hinode Observations of a Vector Magnetic Field
  Change Associated with a Flare on 2006 December 13}}.
\newblock {\it \bibinfo{journal}{Publications of the Astronomical Society of
  Japan}\/},  {\it \bibinfo{volume}{59}\/}, \bibinfo{pages}{S779--S784}.
  \DOIprefix\doi{10.1093/pasj/59.sp3.S779}.
  \href{http://arxiv.org/abs/0709.2397}{\tt arXiv:0709.2397}.
\bibitem[{{K{\"u}nzel}(1960)}]{1960AN....285..271K}
\bibinfo{author}{{K{\"u}nzel}, H.} (\bibinfo{year}{1960}).
\newblock \bibinfo{title}{{Die Flare-H{\"a}ufigkeit in Fleckengruppen
  unterschiedlicher Klasse und magnetischer Struktur (The flare frequency in
  sunspot groups of various classes and magnetic structures)}}.
\newblock {\it \bibinfo{journal}{Astronomische Nachrichten}\/},  {\it
  \bibinfo{volume}{285}\/}\bibinfo{issue}{(5)}, \bibinfo{pages}{271--273}.
  \DOIprefix\doi{10.1002/asna.19592850516}.
\bibitem[{{Kusano} et~al.(2012){Kusano}, {Bamba}, {Yamamoto}, {Iida}, {Toriumi}
  \& {Asai}}]{2012ApJ...760...31K}
\bibinfo{author}{{Kusano}, K.}, \bibinfo{author}{{Bamba}, Y.},
  \bibinfo{author}{{Yamamoto}, T.~T.}, \bibinfo{author}{{Iida}, Y.},
  \bibinfo{author}{{Toriumi}, S.}, \& \bibinfo{author}{{Asai}, A.}
  (\bibinfo{year}{2012}).
\newblock \bibinfo{title}{{Magnetic Field Structures Triggering Solar Flares
  and Coronal Mass Ejections}}.
\newblock {\it \bibinfo{journal}{The Astrophysical Journal}\/},  {\it
  \bibinfo{volume}{760}\/}\bibinfo{issue}{(1)}, \bibinfo{pages}{31}.
  \DOIprefix\doi{10.1088/0004-637X/760/1/31}.
  \href{http://arxiv.org/abs/1210.0598}{\tt arXiv:1210.0598}.
\bibitem[{{Kusano} et~al.(2020){Kusano}, {Iju}, {Bamba} \&
  {Inoue}}]{2020Sci...369..587K}
\bibinfo{author}{{Kusano}, K.}, \bibinfo{author}{{Iju}, T.},
  \bibinfo{author}{{Bamba}, Y.}, \& \bibinfo{author}{{Inoue}, S.}
  (\bibinfo{year}{2020}).
\newblock \bibinfo{title}{{A physics-based method that can predict imminent
  large solar flares}}.
\newblock {\it \bibinfo{journal}{Science}\/},  {\it
  \bibinfo{volume}{369}\/}\bibinfo{issue}{(6503)}, \bibinfo{pages}{587--591}.
  \DOIprefix\doi{10.1126/science.aaz2511}.
\bibitem[{{Leka} \& {Barnes}(2003)}]{2003ApJ...595.1277L}
\bibinfo{author}{{Leka}, K.~D.}, \& \bibinfo{author}{{Barnes}, G.}
  (\bibinfo{year}{2003}).
\newblock \bibinfo{title}{{Photospheric Magnetic Field Properties of Flaring
  versus Flare-quiet Active Regions. I. Data, General Approach, and Sample
  Results}}.
\newblock {\it \bibinfo{journal}{The Astrophysical Journal}\/},  {\it
  \bibinfo{volume}{595}\/}\bibinfo{issue}{(2)}, \bibinfo{pages}{1277--1295}.
  \DOIprefix\doi{10.1086/377511}.
\bibitem[{{Liu} et~al.(2012){Liu}, {Deng}, {Liu}, {Lee}, {Wiegelmann}, {Jing},
  {Xu}, {Wang} \& {Wang}}]{2012ApJ...745L...4L}
\bibinfo{author}{{Liu}, C.}, \bibinfo{author}{{Deng}, N.},
  \bibinfo{author}{{Liu}, R.}, \bibinfo{author}{{Lee}, J.},
  \bibinfo{author}{{Wiegelmann}, T.}, \bibinfo{author}{{Jing}, J.},
  \bibinfo{author}{{Xu}, Y.}, \bibinfo{author}{{Wang}, S.}, \&
  \bibinfo{author}{{Wang}, H.} (\bibinfo{year}{2012}).
\newblock \bibinfo{title}{{Rapid Changes of Photospheric Magnetic Field after
  Tether-cutting Reconnection and Magnetic Implosion}}.
\newblock {\it \bibinfo{journal}{The Astrophysical Journal Letters}\/},  {\it
  \bibinfo{volume}{745}\/}\bibinfo{issue}{(1)}, \bibinfo{pages}{L4}.
  \DOIprefix\doi{10.1088/2041-8205/745/1/L4}.
  \href{http://arxiv.org/abs/1112.3598}{\tt arXiv:1112.3598}.
\bibitem[{{L{\'o}pez Fuentes} et~al.(2000){L{\'o}pez Fuentes}, {Demoulin},
  {Mandrini} \& {van Driel-Gesztelyi}}]{2000ApJ...544..540L}
\bibinfo{author}{{L{\'o}pez Fuentes}, M.~C.}, \bibinfo{author}{{Demoulin}, P.},
  \bibinfo{author}{{Mandrini}, C.~H.}, \& \bibinfo{author}{{van
  Driel-Gesztelyi}, L.} (\bibinfo{year}{2000}).
\newblock \bibinfo{title}{{The Counterkink Rotation of a Non-Hale Active
  Region}}.
\newblock {\it \bibinfo{journal}{The Astrophysical Journal}\/},  {\it
  \bibinfo{volume}{544}\/}\bibinfo{issue}{(1)}, \bibinfo{pages}{540--549}.
  \DOIprefix\doi{10.1086/317180}. \href{http://arxiv.org/abs/1412.1456}{\tt
  arXiv:1412.1456}.
\bibitem[{{Lundquist}(1951)}]{1951PhRv...83..307L}
\bibinfo{author}{{Lundquist}, S.} (\bibinfo{year}{1951}).
\newblock \bibinfo{title}{{On the Stability of Magneto-Hydrostatic Fields}}.
\newblock {\it \bibinfo{journal}{Physical Review}\/},  {\it
  \bibinfo{volume}{83}\/}\bibinfo{issue}{(2)}, \bibinfo{pages}{307--311}.
  \DOIprefix\doi{10.1103/PhysRev.83.307}.
\bibitem[{{Maehara} et~al.(2012){Maehara}, {Shibayama}, {Notsu}, {Notsu},
  {Nagao}, {Kusaba}, {Honda}, {Nogami} \& {Shibata}}]{2012Natur.485..478M}
\bibinfo{author}{{Maehara}, H.}, \bibinfo{author}{{Shibayama}, T.},
  \bibinfo{author}{{Notsu}, S.}, \bibinfo{author}{{Notsu}, Y.},
  \bibinfo{author}{{Nagao}, T.}, \bibinfo{author}{{Kusaba}, S.},
  \bibinfo{author}{{Honda}, S.}, \bibinfo{author}{{Nogami}, D.}, \&
  \bibinfo{author}{{Shibata}, K.} (\bibinfo{year}{2012}).
\newblock \bibinfo{title}{{Superflares on solar-type stars}}.
\newblock {\it \bibinfo{journal}{Nature}\/},  {\it
  \bibinfo{volume}{485}\/}\bibinfo{issue}{(7399)}, \bibinfo{pages}{478--481}.
  \DOIprefix\doi{10.1038/nature11063}.
\bibitem[{{Mandrini} et~al.(2014){Mandrini}, {Schmieder}, {D{\'e}moulin}, {Guo}
  \& {Cristiani}}]{2014SoPh..289.2041M}
\bibinfo{author}{{Mandrini}, C.~H.}, \bibinfo{author}{{Schmieder}, B.},
  \bibinfo{author}{{D{\'e}moulin}, P.}, \bibinfo{author}{{Guo}, Y.}, \&
  \bibinfo{author}{{Cristiani}, G.~D.} (\bibinfo{year}{2014}).
\newblock \bibinfo{title}{{Topological Analysis of Emerging Bipole Clusters
  Producing Violent Solar Events}}.
\newblock {\it \bibinfo{journal}{Solar Physics}\/},  {\it
  \bibinfo{volume}{289}\/}\bibinfo{issue}{(6)}, \bibinfo{pages}{2041--2071}.
  \DOIprefix\doi{10.1007/s11207-013-0458-6}.
  \href{http://arxiv.org/abs/1312.3359}{\tt arXiv:1312.3359}.
\bibitem[{{Min} \& {Chae}(2009)}]{2009SoPh..258..203M}
\bibinfo{author}{{Min}, S.}, \& \bibinfo{author}{{Chae}, J.}
  (\bibinfo{year}{2009}).
\newblock \bibinfo{title}{{The Rotating Sunspot in AR 10930}}.
\newblock {\it \bibinfo{journal}{Solar Physics}\/},  {\it
  \bibinfo{volume}{258}\/}\bibinfo{issue}{(2)}, \bibinfo{pages}{203--217}.
  \DOIprefix\doi{10.1007/s11207-009-9425-7}.
\bibitem[{{Murray} \& {Hood}(2007)}]{2007A&A...470..709M}
\bibinfo{author}{{Murray}, M.~J.}, \& \bibinfo{author}{{Hood}, A.~W.}
  (\bibinfo{year}{2007}).
\newblock \bibinfo{title}{{Simple emergence structures from complex magnetic
  fields}}.
\newblock {\it \bibinfo{journal}{Astronomy and Astrophysics}\/},  {\it
  \bibinfo{volume}{470}\/}\bibinfo{issue}{(2)}, \bibinfo{pages}{709--719}.
  \DOIprefix\doi{10.1051/0004-6361:20077251}.
\bibitem[{{Namekata} et~al.(2019){Namekata}, {Maehara}, {Notsu}, {Toriumi},
  {Hayakawa}, {Ikuta}, {Notsu}, {Honda}, {Nogami} \&
  {Shibata}}]{2019ApJ...871..187N}
\bibinfo{author}{{Namekata}, K.}, \bibinfo{author}{{Maehara}, H.},
  \bibinfo{author}{{Notsu}, Y.}, \bibinfo{author}{{Toriumi}, S.},
  \bibinfo{author}{{Hayakawa}, H.}, \bibinfo{author}{{Ikuta}, K.},
  \bibinfo{author}{{Notsu}, S.}, \bibinfo{author}{{Honda}, S.},
  \bibinfo{author}{{Nogami}, D.}, \& \bibinfo{author}{{Shibata}, K.}
  (\bibinfo{year}{2019}).
\newblock \bibinfo{title}{{Lifetimes and Emergence/Decay Rates of Star Spots on
  Solar-type Stars Estimated by Kepler Data in Comparison with Those of
  Sunspots}}.
\newblock {\it \bibinfo{journal}{The Astrophysical Journal}\/},  {\it
  \bibinfo{volume}{871}\/}\bibinfo{issue}{(2)}, \bibinfo{pages}{187}.
  \DOIprefix\doi{10.3847/1538-4357/aaf471}.
  \href{http://arxiv.org/abs/1811.10782}{\tt arXiv:1811.10782}.
\bibitem[{{Nelson} et~al.(2013){Nelson}, {Brown}, {Brun}, {Miesch} \&
  {Toomre}}]{2013ApJ...762...73N}
\bibinfo{author}{{Nelson}, N.~J.}, \bibinfo{author}{{Brown}, B.~P.},
  \bibinfo{author}{{Brun}, A.~S.}, \bibinfo{author}{{Miesch}, M.~S.}, \&
  \bibinfo{author}{{Toomre}, J.} (\bibinfo{year}{2013}).
\newblock \bibinfo{title}{{Magnetic Wreaths and Cycles in Convective Dynamos}}.
\newblock {\it \bibinfo{journal}{The Astrophysical Journal}\/},  {\it
  \bibinfo{volume}{762}\/}\bibinfo{issue}{(2)}, \bibinfo{pages}{73}.
  \DOIprefix\doi{10.1088/0004-637X/762/2/73}.
  \href{http://arxiv.org/abs/1211.3129}{\tt arXiv:1211.3129}.
\bibitem[{{Okamoto} \& {Sakurai}(2018)}]{2018ApJ...852L..16O}
\bibinfo{author}{{Okamoto}, T.~J.}, \& \bibinfo{author}{{Sakurai}, T.}
  (\bibinfo{year}{2018}).
\newblock \bibinfo{title}{{Super-strong Magnetic Field in Sunspots}}.
\newblock {\it \bibinfo{journal}{The Astrophysical Journal Letters}\/},  {\it
  \bibinfo{volume}{852}\/}\bibinfo{issue}{(1)}, \bibinfo{pages}{L16}.
  \DOIprefix\doi{10.3847/2041-8213/aaa3d8}.
  \href{http://arxiv.org/abs/1712.08700}{\tt arXiv:1712.08700}.
\bibitem[{{Parker}(1955)}]{1955ApJ...121..491P}
\bibinfo{author}{{Parker}, E.~N.} (\bibinfo{year}{1955}).
\newblock \bibinfo{title}{{The Formation of Sunspots from the Solar Toroidal
  Field.}}
\newblock {\it \bibinfo{journal}{Astrophysical Journal}\/},  {\it
  \bibinfo{volume}{121}\/}, \bibinfo{pages}{491--507}.
  \DOIprefix\doi{10.1086/146010}.
\bibitem[{{Parker}(1979)}]{1979cmft.book.....P}
\bibinfo{author}{{Parker}, E.~N.} (\bibinfo{year}{1979}).
\newblock {\it \bibinfo{title}{{Cosmical magnetic fields. Their origin and
  their activity}}\/}.
\newblock \bibinfo{address}{Oxford}: \bibinfo{publisher}{Clarendon Press}.
\bibitem[{{Pevtsov}(2002)}]{2002SoPh..207..111P}
\bibinfo{author}{{Pevtsov}, A.~A.} (\bibinfo{year}{2002}).
\newblock \bibinfo{title}{{Active-Region Filaments and X-ray Sigmoids}}.
\newblock {\it \bibinfo{journal}{Solar Physics}\/},  {\it
  \bibinfo{volume}{207}\/}\bibinfo{issue}{(1)}, \bibinfo{pages}{111--123}.
  \DOIprefix\doi{10.1023/A:1015589802234}.
\bibitem[{{Poisson} et~al.(2015){Poisson}, {Mandrini}, {D{\'e}moulin} \&
  {L{\'o}pez Fuentes}}]{2015SoPh..290..727P}
\bibinfo{author}{{Poisson}, M.}, \bibinfo{author}{{Mandrini}, C.~H.},
  \bibinfo{author}{{D{\'e}moulin}, P.}, \& \bibinfo{author}{{L{\'o}pez
  Fuentes}, M.} (\bibinfo{year}{2015}).
\newblock \bibinfo{title}{{Evidence of Twisted Flux-Tube Emergence in Active
  Regions}}.
\newblock {\it \bibinfo{journal}{Solar Physics}\/},  {\it
  \bibinfo{volume}{290}\/}\bibinfo{issue}{(3)}, \bibinfo{pages}{727--751}.
  \DOIprefix\doi{10.1007/s11207-014-0633-4}.
  \href{http://arxiv.org/abs/1505.01805}{\tt arXiv:1505.01805}.
\bibitem[{{Priest} \& {Forbes}(2002)}]{2002A&ARv..10..313P}
\bibinfo{author}{{Priest}, E.~R.}, \& \bibinfo{author}{{Forbes}, T.~G.}
  (\bibinfo{year}{2002}).
\newblock \bibinfo{title}{{The magnetic nature of solar flares}}.
\newblock {\it \bibinfo{journal}{The Astronomy and Astrophysics Review}\/},
  {\it \bibinfo{volume}{10}\/}\bibinfo{issue}{(4)}, \bibinfo{pages}{313--377}.
  \DOIprefix\doi{10.1007/s001590100013}.
\bibitem[{{Rempel}(2005)}]{2005ApJ...622.1320R}
\bibinfo{author}{{Rempel}, M.} (\bibinfo{year}{2005}).
\newblock \bibinfo{title}{{Solar Differential Rotation and Meridional Flow: The
  Role of a Subadiabatic Tachocline for the Taylor-Proudman Balance}}.
\newblock {\it \bibinfo{journal}{The Astrophysical Journal}\/},  {\it
  \bibinfo{volume}{622}\/}\bibinfo{issue}{(2)}, \bibinfo{pages}{1320--1332}.
  \DOIprefix\doi{10.1086/428282}.
  \href{http://arxiv.org/abs/astro-ph/0604451}{\tt arXiv:astro-ph/0604451}.
\bibitem[{{Rempel} \& {Cheung}(2014)}]{2014ApJ...785...90R}
\bibinfo{author}{{Rempel}, M.}, \& \bibinfo{author}{{Cheung}, M.~C.~M.}
  (\bibinfo{year}{2014}).
\newblock \bibinfo{title}{{Numerical Simulations of Active Region Scale Flux
  Emergence: From Spot Formation to Decay}}.
\newblock {\it \bibinfo{journal}{The Astrophysical Journal}\/},  {\it
  \bibinfo{volume}{785}\/}\bibinfo{issue}{(2)}, \bibinfo{pages}{90}.
  \DOIprefix\doi{10.1088/0004-637X/785/2/90}.
  \href{http://arxiv.org/abs/1402.4703}{\tt arXiv:1402.4703}.
\bibitem[{{Rimmele} et~al.(2020){Rimmele}, {Warner}, {Keil}, {Goode},
  {Kn{\"o}lker}, {Kuhn}, {Rosner}, {McMullin}, {Casini}, {Lin}, {W{\"o}ger},
  {von der L{\"u}he}, {Tritschler}, {Davey}, {de Wijn}, {Elmore}, {Fehlmann},
  {Harrington}, {Jaeggli}, {Rast}, {Schad}, {Schmidt}, {Mathioudakis},
  {Mickey}, {Anan}, {Beck}, {Marshall}, {Jeffers}, {Oschmann}, {Beard},
  {Berst}, {Cowan}, {Craig}, {Cross}, {Cummings}, {Donnelly}, {de Vanssay},
  {Eigenbrot}, {Ferayorni}, {Foster}, {Galapon}, {Gedrites}, {Gonzales},
  {Goodrich}, {Gregory}, {Guzman}, {Guzzo}, {Hegwer}, {Hubbard}, {Hubbard},
  {Johansson}, {Johnson}, {Liang}, {Liang}, {McQuillen}, {Mayer}, {Newman},
  {Onodera}, {Phelps}, {Puentes}, {Richards}, {Rimmele}, {Sekulic}, {Shimko},
  {Simison}, {Smith}, {Starman}, {Sueoka}, {Summers}, {Szabo}, {Szabo},
  {Wampler}, {Williams} \& {White}}]{2020SoPh..295..172R}
\bibinfo{author}{{Rimmele}, T.~R.}, \bibinfo{author}{{Warner}, M.},
  \bibinfo{author}{{Keil}, S.~L.}, \bibinfo{author}{{Goode}, P.~R.},
  \bibinfo{author}{{Kn{\"o}lker}, M.}, \bibinfo{author}{{Kuhn}, J.~R.},
  \bibinfo{author}{{Rosner}, R.~R.}, \bibinfo{author}{{McMullin}, J.~P.},
  \bibinfo{author}{{Casini}, R.}, \bibinfo{author}{{Lin}, H.},
  \bibinfo{author}{{W{\"o}ger}, F.}, \bibinfo{author}{{von der L{\"u}he}, O.},
  \bibinfo{author}{{Tritschler}, A.}, \bibinfo{author}{{Davey}, A.},
  \bibinfo{author}{{de Wijn}, A.}, \bibinfo{author}{{Elmore}, D.~F.},
  \bibinfo{author}{{Fehlmann}, A.}, \bibinfo{author}{{Harrington}, D.~M.},
  \bibinfo{author}{{Jaeggli}, S.~A.}, \bibinfo{author}{{Rast}, M.~P.},
  \bibinfo{author}{{Schad}, T.~A.}, \bibinfo{author}{{Schmidt}, W.},
  \bibinfo{author}{{Mathioudakis}, M.}, \bibinfo{author}{{Mickey}, D.~L.},
  \bibinfo{author}{{Anan}, T.}, \bibinfo{author}{{Beck}, C.},
  \bibinfo{author}{{Marshall}, H.~K.}, \bibinfo{author}{{Jeffers}, P.~F.},
  \bibinfo{author}{{Oschmann}, J.~M.}, \bibinfo{author}{{Beard}, A.},
  \bibinfo{author}{{Berst}, D.~C.}, \bibinfo{author}{{Cowan}, B.~A.},
  \bibinfo{author}{{Craig}, S.~C.}, \bibinfo{author}{{Cross}, E.},
  \bibinfo{author}{{Cummings}, B.~K.}, \bibinfo{author}{{Donnelly}, C.},
  \bibinfo{author}{{de Vanssay}, J.-B.}, \bibinfo{author}{{Eigenbrot}, A.~D.},
  \bibinfo{author}{{Ferayorni}, A.}, \bibinfo{author}{{Foster}, C.},
  \bibinfo{author}{{Galapon}, C.~A.}, \bibinfo{author}{{Gedrites}, C.},
  \bibinfo{author}{{Gonzales}, K.}, \bibinfo{author}{{Goodrich}, B.~D.},
  \bibinfo{author}{{Gregory}, B.~S.}, \bibinfo{author}{{Guzman}, S.~S.},
  \bibinfo{author}{{Guzzo}, S.}, \bibinfo{author}{{Hegwer}, S.},
  \bibinfo{author}{{Hubbard}, R.~P.}, \bibinfo{author}{{Hubbard}, J.~R.},
  \bibinfo{author}{{Johansson}, E.~M.}, \bibinfo{author}{{Johnson}, L.~C.},
  \bibinfo{author}{{Liang}, C.}, \bibinfo{author}{{Liang}, M.},
  \bibinfo{author}{{McQuillen}, I.}, \bibinfo{author}{{Mayer}, C.},
  \bibinfo{author}{{Newman}, K.}, \bibinfo{author}{{Onodera}, B.},
  \bibinfo{author}{{Phelps}, L.}, \bibinfo{author}{{Puentes}, M.~M.},
  \bibinfo{author}{{Richards}, C.}, \bibinfo{author}{{Rimmele}, L.~M.},
  \bibinfo{author}{{Sekulic}, P.}, \bibinfo{author}{{Shimko}, S.~R.},
  \bibinfo{author}{{Simison}, B.~E.}, \bibinfo{author}{{Smith}, B.},
  \bibinfo{author}{{Starman}, E.}, \bibinfo{author}{{Sueoka}, S.~R.},
  \bibinfo{author}{{Summers}, R.~T.}, \bibinfo{author}{{Szabo}, A.},
  \bibinfo{author}{{Szabo}, L.}, \bibinfo{author}{{Wampler}, S.~B.},
  \bibinfo{author}{{Williams}, T.~R.}, \& \bibinfo{author}{{White}, C.}
  (\bibinfo{year}{2020}).
\newblock \bibinfo{title}{{The Daniel K. Inouye Solar Telescope - Observatory
  Overview}}.
\newblock {\it \bibinfo{journal}{Solar Physics}\/},  {\it
  \bibinfo{volume}{295}\/}\bibinfo{issue}{(12)}, \bibinfo{pages}{172}.
  \DOIprefix\doi{10.1007/s11207-020-01736-7}.
\bibitem[{{Rust} \& {Kumar}(1996)}]{1996ApJ...464L.199R}
\bibinfo{author}{{Rust}, D.~M.}, \& \bibinfo{author}{{Kumar}, A.}
  (\bibinfo{year}{1996}).
\newblock \bibinfo{title}{{Evidence for Helically Kinked Magnetic Flux Ropes in
  Solar Eruptions}}.
\newblock {\it \bibinfo{journal}{Astrophysical Journal Letters}\/},  {\it
  \bibinfo{volume}{464}\/}, \bibinfo{pages}{L199--L202}.
  \DOIprefix\doi{10.1086/310118}.
\bibitem[{{Sammis} et~al.(2000){Sammis}, {Tang} \&
  {Zirin}}]{2000ApJ...540..583S}
\bibinfo{author}{{Sammis}, I.}, \bibinfo{author}{{Tang}, F.}, \&
  \bibinfo{author}{{Zirin}, H.} (\bibinfo{year}{2000}).
\newblock \bibinfo{title}{{The Dependence of Large Flare Occurrence on the
  Magnetic Structure of Sunspots}}.
\newblock {\it \bibinfo{journal}{The Astrophysical Journal}\/},  {\it
  \bibinfo{volume}{540}\/}\bibinfo{issue}{(1)}, \bibinfo{pages}{583--587}.
  \DOIprefix\doi{10.1086/309303}.
\bibitem[{{Schmieder} et~al.(2014){Schmieder}, {Archontis} \&
  {Pariat}}]{2014SSRv..186..227S}
\bibinfo{author}{{Schmieder}, B.}, \bibinfo{author}{{Archontis}, V.}, \&
  \bibinfo{author}{{Pariat}, E.} (\bibinfo{year}{2014}).
\newblock \bibinfo{title}{{Magnetic Flux Emergence Along the Solar Cycle}}.
\newblock {\it \bibinfo{journal}{Space Science Reviews}\/},  {\it
  \bibinfo{volume}{186}\/}\bibinfo{issue}{(1-4)}, \bibinfo{pages}{227--250}.
  \DOIprefix\doi{10.1007/s11214-014-0088-9}.
\bibitem[{{Schmieder} et~al.(1994){Schmieder}, {Hagyard}, {Guoxiang}, {Hongqi},
  {Kalman}, {Gyori}, {Rompolt}, {Demoulin} \& {Machado}}]{1994SoPh..150..199S}
\bibinfo{author}{{Schmieder}, B.}, \bibinfo{author}{{Hagyard}, M.~J.},
  \bibinfo{author}{{Guoxiang}, A.}, \bibinfo{author}{{Hongqi}, Z.},
  \bibinfo{author}{{Kalman}, B.}, \bibinfo{author}{{Gyori}, L.},
  \bibinfo{author}{{Rompolt}, B.}, \bibinfo{author}{{Demoulin}, P.}, \&
  \bibinfo{author}{{Machado}, M.~E.} (\bibinfo{year}{1994}).
\newblock \bibinfo{title}{{Relationship between magnetic field evolution and
  flaring sites in AR 6659 in June 1991}}.
\newblock {\it \bibinfo{journal}{Solar Physics}\/},  {\it
  \bibinfo{volume}{150}\/}\bibinfo{issue}{(1-2)}, \bibinfo{pages}{199--219}.
  \DOIprefix\doi{10.1007/BF00712886}.
\bibitem[{{Schrijver}(2007)}]{2007ApJ...655L.117S}
\bibinfo{author}{{Schrijver}, C.~J.} (\bibinfo{year}{2007}).
\newblock \bibinfo{title}{{A Characteristic Magnetic Field Pattern Associated
  with All Major Solar Flares and Its Use in Flare Forecasting}}.
\newblock {\it \bibinfo{journal}{The Astrophysical Journal Letters}\/},  {\it
  \bibinfo{volume}{655}\/}\bibinfo{issue}{(2)}, \bibinfo{pages}{L117--L120}.
  \DOIprefix\doi{10.1086/511857}.
\bibitem[{{Schrijver} et~al.(2015){Schrijver}, {Kauristie}, {Aylward},
  {Denardini}, {Gibson}, {Glover}, {Gopalswamy}, {Grande}, {Hapgood},
  {Heynderickx}, {Jakowski}, {Kalegaev}, {Lapenta}, {Linker}, {Liu},
  {Mandrini}, {Mann}, {Nagatsuma}, {Nandy}, {Obara}, {Paul O'Brien}, {Onsager},
  {Opgenoorth}, {Terkildsen}, {Valladares} \& {Vilmer}}]{2015AdSpR..55.2745S}
\bibinfo{author}{{Schrijver}, C.~J.}, \bibinfo{author}{{Kauristie}, K.},
  \bibinfo{author}{{Aylward}, A.~D.}, \bibinfo{author}{{Denardini}, C.~M.},
  \bibinfo{author}{{Gibson}, S.~E.}, \bibinfo{author}{{Glover}, A.},
  \bibinfo{author}{{Gopalswamy}, N.}, \bibinfo{author}{{Grande}, M.},
  \bibinfo{author}{{Hapgood}, M.}, \bibinfo{author}{{Heynderickx}, D.},
  \bibinfo{author}{{Jakowski}, N.}, \bibinfo{author}{{Kalegaev}, V.~V.},
  \bibinfo{author}{{Lapenta}, G.}, \bibinfo{author}{{Linker}, J.~A.},
  \bibinfo{author}{{Liu}, S.}, \bibinfo{author}{{Mandrini}, C.~H.},
  \bibinfo{author}{{Mann}, I.~R.}, \bibinfo{author}{{Nagatsuma}, T.},
  \bibinfo{author}{{Nandy}, D.}, \bibinfo{author}{{Obara}, T.},
  \bibinfo{author}{{Paul O'Brien}, T.}, \bibinfo{author}{{Onsager}, T.},
  \bibinfo{author}{{Opgenoorth}, H.~J.}, \bibinfo{author}{{Terkildsen}, M.},
  \bibinfo{author}{{Valladares}, C.~E.}, \& \bibinfo{author}{{Vilmer}, N.}
  (\bibinfo{year}{2015}).
\newblock \bibinfo{title}{{Understanding space weather to shield society: A
  global road map for 2015-2025 commissioned by COSPAR and ILWS}}.
\newblock {\it \bibinfo{journal}{Advances in Space Research}\/},  {\it
  \bibinfo{volume}{55}\/}\bibinfo{issue}{(12)}, \bibinfo{pages}{2745--2807}.
  \DOIprefix\doi{10.1016/j.asr.2015.03.023}.
  \href{http://arxiv.org/abs/1503.06135}{\tt arXiv:1503.06135}.
\bibitem[{{Severnyj}(1958)}]{1958IzKry..20...22S}
\bibinfo{author}{{Severnyj}, A.~B.} (\bibinfo{year}{1958}).
\newblock \bibinfo{title}{{The Appearance of Flares in Neutral Points of the
  Solar Magnetic Field and the Pinch-effect}}.
\newblock {\it \bibinfo{journal}{Izvestiya Ordena Trudovogo Krasnogo Znameni
  Krymskoj Astrofizicheskoj Observatorii (Bulletin of the Crimean Astrophysical
  Observatory)}\/},  {\it \bibinfo{volume}{20}\/}, \bibinfo{pages}{22--51}.
\bibitem[{{Shibata} et~al.(2013){Shibata}, {Isobe}, {Hillier}, {Choudhuri},
  {Maehara}, {Ishii}, {Shibayama}, {Notsu}, {Notsu}, {Nagao}, {Honda} \&
  {Nogami}}]{2013PASJ...65...49S}
\bibinfo{author}{{Shibata}, K.}, \bibinfo{author}{{Isobe}, H.},
  \bibinfo{author}{{Hillier}, A.}, \bibinfo{author}{{Choudhuri}, A.~R.},
  \bibinfo{author}{{Maehara}, H.}, \bibinfo{author}{{Ishii}, T.~T.},
  \bibinfo{author}{{Shibayama}, T.}, \bibinfo{author}{{Notsu}, S.},
  \bibinfo{author}{{Notsu}, Y.}, \bibinfo{author}{{Nagao}, T.},
  \bibinfo{author}{{Honda}, S.}, \& \bibinfo{author}{{Nogami}, D.}
  (\bibinfo{year}{2013}).
\newblock \bibinfo{title}{{Can Superflares Occur on Our Sun?}}
\newblock {\it \bibinfo{journal}{Publications of the Astronomical Society of
  Japan}\/},  {\it \bibinfo{volume}{65}\/}, \bibinfo{pages}{49}.
  \DOIprefix\doi{10.1093/pasj/65.3.49}.
  \href{http://arxiv.org/abs/1212.1361}{\tt arXiv:1212.1361}.
\bibitem[{{Shibata} \& {Magara}(2011)}]{2011LRSP....8....6S}
\bibinfo{author}{{Shibata}, K.}, \& \bibinfo{author}{{Magara}, T.}
  (\bibinfo{year}{2011}).
\newblock \bibinfo{title}{{Solar Flares: Magnetohydrodynamic Processes}}.
\newblock {\it \bibinfo{journal}{Living Reviews in Solar Physics}\/},  {\it
  \bibinfo{volume}{8}\/}\bibinfo{issue}{(1)}, \bibinfo{pages}{6}.
  \DOIprefix\doi{10.12942/lrsp-2011-6}.
\bibitem[{{Shimizu} et~al.(2019){Shimizu}, {Imada}, {Kawate}, {Ichimoto},
  {Suematsu}, {Hara}, {Katsukawa}, {Kubo}, {Toriumi}, {Watanabe}, {Yokoyama},
  {Korendyke}, {Warren}, {Tarbell}, {De Pontieu}, {Teriaca}, {Sch{\"u}hle},
  {Solanki}, {Harra}, {Matthews}, {Fludra}, {Auch{\`e}re}, {Andretta},
  {Naletto} \& {Zhukov}}]{2019SPIE11118E..07S}
\bibinfo{author}{{Shimizu}, T.}, \bibinfo{author}{{Imada}, S.},
  \bibinfo{author}{{Kawate}, T.}, \bibinfo{author}{{Ichimoto}, K.},
  \bibinfo{author}{{Suematsu}, Y.}, \bibinfo{author}{{Hara}, H.},
  \bibinfo{author}{{Katsukawa}, Y.}, \bibinfo{author}{{Kubo}, M.},
  \bibinfo{author}{{Toriumi}, S.}, \bibinfo{author}{{Watanabe}, T.},
  \bibinfo{author}{{Yokoyama}, T.}, \bibinfo{author}{{Korendyke}, C.~M.},
  \bibinfo{author}{{Warren}, H.~P.}, \bibinfo{author}{{Tarbell}, T.},
  \bibinfo{author}{{De Pontieu}, B.}, \bibinfo{author}{{Teriaca}, L.},
  \bibinfo{author}{{Sch{\"u}hle}, U.~H.}, \bibinfo{author}{{Solanki}, S.},
  \bibinfo{author}{{Harra}, L.~K.}, \bibinfo{author}{{Matthews}, S.},
  \bibinfo{author}{{Fludra}, A.}, \bibinfo{author}{{Auch{\`e}re}, F.},
  \bibinfo{author}{{Andretta}, V.}, \bibinfo{author}{{Naletto}, G.}, \&
  \bibinfo{author}{{Zhukov}, A.} (\bibinfo{year}{2019}).
\newblock \bibinfo{title}{{The Solar-C\_EUVST mission}}.
\newblock In {\it \bibinfo{booktitle}{UV, X-Ray, and Gamma-Ray Space
  Instrumentation for Astronomy XXI}\/} (p. \bibinfo{pages}{1111807}).
\newblock volume \bibinfo{volume}{11118} of {\it \bibinfo{series}{Society of
  Photo-Optical Instrumentation Engineers (SPIE) Conference Series}\/}.
\newblock \DOIprefix\doi{10.1117/12.2528240}.
\bibitem[{{Sturrock}(1966)}]{1966Natur.211..695S}
\bibinfo{author}{{Sturrock}, P.~A.} (\bibinfo{year}{1966}).
\newblock \bibinfo{title}{{Model of the High-Energy Phase of Solar Flares}}.
\newblock {\it \bibinfo{journal}{Nature}\/},  {\it
  \bibinfo{volume}{211}\/}\bibinfo{issue}{(5050)}, \bibinfo{pages}{695--697}.
  \DOIprefix\doi{10.1038/211695a0}.
\bibitem[{{Syntelis} et~al.(2019){Syntelis}, {Lee}, {Fairbairn}, {Archontis} \&
  {Hood}}]{2019A&A...630A.134S}
\bibinfo{author}{{Syntelis}, P.}, \bibinfo{author}{{Lee}, E.~J.},
  \bibinfo{author}{{Fairbairn}, C.~W.}, \bibinfo{author}{{Archontis}, V.}, \&
  \bibinfo{author}{{Hood}, A.~W.} (\bibinfo{year}{2019}).
\newblock \bibinfo{title}{{Eruptions and flaring activity in emerging
  quadrupolar regions}}.
\newblock {\it \bibinfo{journal}{Astronomy \& Astrophysics}\/},  {\it
  \bibinfo{volume}{630}\/}, \bibinfo{pages}{A134}.
  \DOIprefix\doi{10.1051/0004-6361/201936246}.
  \href{http://arxiv.org/abs/1909.01446}{\tt arXiv:1909.01446}.
\bibitem[{{Takasao} et~al.(2015){Takasao}, {Fan}, {Cheung} \&
  {Shibata}}]{2015ApJ...813..112T}
\bibinfo{author}{{Takasao}, S.}, \bibinfo{author}{{Fan}, Y.},
  \bibinfo{author}{{Cheung}, M. C.~M.}, \& \bibinfo{author}{{Shibata}, K.}
  (\bibinfo{year}{2015}).
\newblock \bibinfo{title}{{Numerical Study on the Emergence of Kinked Flux Tube
  for Understanding of Possible Origin of {\ensuremath{\delta}}-spot Regions}}.
\newblock {\it \bibinfo{journal}{The Astrophysical Journal}\/},  {\it
  \bibinfo{volume}{813}\/}\bibinfo{issue}{(2)}, \bibinfo{pages}{112}.
  \DOIprefix\doi{10.1088/0004-637X/813/2/112}.
  \href{http://arxiv.org/abs/1511.02863}{\tt arXiv:1511.02863}.
\bibitem[{{Tanaka}(1991)}]{1991SoPh..136..133T}
\bibinfo{author}{{Tanaka}, K.} (\bibinfo{year}{1991}).
\newblock \bibinfo{title}{{Studies on a very flare-active {\ensuremath{\delta}}
  group: Peculiar {\ensuremath{\delta}} spot evolution and inferred subsurface
  magnetic rope structure}}.
\newblock {\it \bibinfo{journal}{Solar Physics}\/},  {\it
  \bibinfo{volume}{136}\/}\bibinfo{issue}{(1)}, \bibinfo{pages}{133--149}.
  \DOIprefix\doi{10.1007/BF00151700}.
\bibitem[{{Tang} \& {Wang}(1993)}]{1993SoPh..143..107T}
\bibinfo{author}{{Tang}, F.}, \& \bibinfo{author}{{Wang}, H.}
  (\bibinfo{year}{1993}).
\newblock \bibinfo{title}{{On the Dynamic Activity in Sheared Corridors of
  Large Delta Spots}}.
\newblock {\it \bibinfo{journal}{Solar Physics}\/},  {\it
  \bibinfo{volume}{143}\/}\bibinfo{issue}{(1)}, \bibinfo{pages}{107--118}.
  \DOIprefix\doi{10.1007/BF00619099}.
\bibitem[{{Toriumi} et~al.(2020){Toriumi}, {Airapetian}, {Hudson}, {Schrijver},
  {Cheung} \& {DeRosa}}]{2020ApJ...902...36T}
\bibinfo{author}{{Toriumi}, S.}, \bibinfo{author}{{Airapetian}, V.~S.},
  \bibinfo{author}{{Hudson}, H.~S.}, \bibinfo{author}{{Schrijver}, C.~J.},
  \bibinfo{author}{{Cheung}, M. C.~M.}, \& \bibinfo{author}{{DeRosa}, M.~L.}
  (\bibinfo{year}{2020}).
\newblock \bibinfo{title}{{Sun-as-a-star Spectral Irradiance Observations of
  Transiting Active Regions}}.
\newblock {\it \bibinfo{journal}{The Astrophysical Journal}\/},  {\it
  \bibinfo{volume}{902}\/}\bibinfo{issue}{(1)}, \bibinfo{pages}{36}.
  \DOIprefix\doi{10.3847/1538-4357/abadf9}.
  \href{http://arxiv.org/abs/2008.04319}{\tt arXiv:2008.04319}.
\bibitem[{{Toriumi} et~al.(2015{\natexlab{a}}){Toriumi}, {Cheung} \&
  {Katsukawa}}]{2015ApJ...811..138T}
\bibinfo{author}{{Toriumi}, S.}, \bibinfo{author}{{Cheung}, M. C.~M.}, \&
  \bibinfo{author}{{Katsukawa}, Y.} (\bibinfo{year}{2015}{\natexlab{a}}).
\newblock \bibinfo{title}{{Light Bridge in a Developing Active Region. II.
  Numerical Simulation of Flux Emergence and Light Bridge Formation}}.
\newblock {\it \bibinfo{journal}{The Astrophysical Journal}\/},  {\it
  \bibinfo{volume}{811}\/}\bibinfo{issue}{(2)}, \bibinfo{pages}{138}.
  \DOIprefix\doi{10.1088/0004-637X/811/2/138}.
  \href{http://arxiv.org/abs/1509.00205}{\tt arXiv:1509.00205}.
\bibitem[{{Toriumi} \& {Hotta}(2019)}]{2019ApJ...886L..21T}
\bibinfo{author}{{Toriumi}, S.}, \& \bibinfo{author}{{Hotta}, H.}
  (\bibinfo{year}{2019}).
\newblock \bibinfo{title}{{Spontaneous Generation of
  {\ensuremath{\delta}}-sunspots in Convective Magnetohydrodynamic Simulation
  of Magnetic Flux Emergence}}.
\newblock {\it \bibinfo{journal}{The Astrophysical Journal Letters}\/},  {\it
  \bibinfo{volume}{886}\/}\bibinfo{issue}{(1)}, \bibinfo{pages}{L21}.
  \DOIprefix\doi{10.3847/2041-8213/ab55e7}.
  \href{http://arxiv.org/abs/1911.03909}{\tt arXiv:1911.03909}.
\bibitem[{{Toriumi} et~al.(2013){Toriumi}, {Iida}, {Bamba}, {Kusano}, {Imada}
  \& {Inoue}}]{2013ApJ...773..128T}
\bibinfo{author}{{Toriumi}, S.}, \bibinfo{author}{{Iida}, Y.},
  \bibinfo{author}{{Bamba}, Y.}, \bibinfo{author}{{Kusano}, K.},
  \bibinfo{author}{{Imada}, S.}, \& \bibinfo{author}{{Inoue}, S.}
  (\bibinfo{year}{2013}).
\newblock \bibinfo{title}{{The Magnetic Systems Triggering the M6.6 Class Solar
  Flare in NOAA Active Region 11158}}.
\newblock {\it \bibinfo{journal}{The Astrophysical Journal}\/},  {\it
  \bibinfo{volume}{773}\/}\bibinfo{issue}{(2)}, \bibinfo{pages}{128}.
  \DOIprefix\doi{10.1088/0004-637X/773/2/128}.
  \href{http://arxiv.org/abs/1306.2451}{\tt arXiv:1306.2451}.
\bibitem[{{Toriumi} et~al.(2014){Toriumi}, {Iida}, {Kusano}, {Bamba} \&
  {Imada}}]{2014SoPh..289.3351T}
\bibinfo{author}{{Toriumi}, S.}, \bibinfo{author}{{Iida}, Y.},
  \bibinfo{author}{{Kusano}, K.}, \bibinfo{author}{{Bamba}, Y.}, \&
  \bibinfo{author}{{Imada}, S.} (\bibinfo{year}{2014}).
\newblock \bibinfo{title}{{Formation of a Flare-Productive Active Region:
  Observation and Numerical Simulation of NOAA AR 11158}}.
\newblock {\it \bibinfo{journal}{Solar Physics}\/},  {\it
  \bibinfo{volume}{289}\/}\bibinfo{issue}{(9)}, \bibinfo{pages}{3351--3369}.
  \DOIprefix\doi{10.1007/s11207-014-0502-1}.
  \href{http://arxiv.org/abs/1403.4029}{\tt arXiv:1403.4029}.
\bibitem[{{Toriumi} et~al.(2015{\natexlab{b}}){Toriumi}, {Katsukawa} \&
  {Cheung}}]{2015ApJ...811..137T}
\bibinfo{author}{{Toriumi}, S.}, \bibinfo{author}{{Katsukawa}, Y.}, \&
  \bibinfo{author}{{Cheung}, M. C.~M.} (\bibinfo{year}{2015}{\natexlab{b}}).
\newblock \bibinfo{title}{{Light Bridge in a Developing Active Region. I.
  Observation of Light Bridge and its Dynamic Activity Phenomena}}.
\newblock {\it \bibinfo{journal}{The Astrophysical Journal}\/},  {\it
  \bibinfo{volume}{811}\/}\bibinfo{issue}{(2)}, \bibinfo{pages}{137}.
  \DOIprefix\doi{10.1088/0004-637X/811/2/137}.
  \href{http://arxiv.org/abs/1509.00183}{\tt arXiv:1509.00183}.
\bibitem[{{Toriumi} et~al.(2017){Toriumi}, {Schrijver}, {Harra}, {Hudson} \&
  {Nagashima}}]{2017ApJ...834...56T}
\bibinfo{author}{{Toriumi}, S.}, \bibinfo{author}{{Schrijver}, C.~J.},
  \bibinfo{author}{{Harra}, L.~K.}, \bibinfo{author}{{Hudson}, H.}, \&
  \bibinfo{author}{{Nagashima}, K.} (\bibinfo{year}{2017}).
\newblock \bibinfo{title}{{Magnetic Properties of Solar Active Regions That
  Govern Large Solar Flares and Eruptions}}.
\newblock {\it \bibinfo{journal}{The Astrophysical Journal}\/},  {\it
  \bibinfo{volume}{834}\/}\bibinfo{issue}{(1)}, \bibinfo{pages}{56}.
  \DOIprefix\doi{10.3847/1538-4357/834/1/56}.
  \href{http://arxiv.org/abs/1611.05047}{\tt arXiv:1611.05047}.
\bibitem[{{Toriumi} \& {Takasao}(2017)}]{2017ApJ...850...39T}
\bibinfo{author}{{Toriumi}, S.}, \& \bibinfo{author}{{Takasao}, S.}
  (\bibinfo{year}{2017}).
\newblock \bibinfo{title}{{Numerical Simulations of Flare-productive Active
  Regions: {\ensuremath{\delta}}-sunspots, Sheared Polarity Inversion Lines,
  Energy Storage, and Predictions}}.
\newblock {\it \bibinfo{journal}{The Astrophysical Journal}\/},  {\it
  \bibinfo{volume}{850}\/}\bibinfo{issue}{(1)}, \bibinfo{pages}{39}.
  \DOIprefix\doi{10.3847/1538-4357/aa95c2}.
  \href{http://arxiv.org/abs/1710.08926}{\tt arXiv:1710.08926}.
\bibitem[{{Toriumi} \& {Wang}(2019)}]{2019LRSP...16....3T}
\bibinfo{author}{{Toriumi}, S.}, \& \bibinfo{author}{{Wang}, H.}
  (\bibinfo{year}{2019}).
\newblock \bibinfo{title}{{Flare-productive active regions}}.
\newblock {\it \bibinfo{journal}{Living Reviews in Solar Physics}\/},  {\it
  \bibinfo{volume}{16}\/}\bibinfo{issue}{(1)}, \bibinfo{pages}{3}.
  \DOIprefix\doi{10.1007/s41116-019-0019-7}.
  \href{http://arxiv.org/abs/1904.12027}{\tt arXiv:1904.12027}.
\bibitem[{{Tsuneta} et~al.(2008){Tsuneta}, {Ichimoto}, {Katsukawa}, {Nagata},
  {Otsubo}, {Shimizu}, {Suematsu}, {Nakagiri}, {Noguchi}, {Tarbell}, {Title},
  {Shine}, {Rosenberg}, {Hoffmann}, {Jurcevich}, {Kushner}, {Levay}, {Lites},
  {Elmore}, {Matsushita}, {Kawaguchi}, {Saito}, {Mikami}, {Hill} \&
  {Owens}}]{2008SoPh..249..167T}
\bibinfo{author}{{Tsuneta}, S.}, \bibinfo{author}{{Ichimoto}, K.},
  \bibinfo{author}{{Katsukawa}, Y.}, \bibinfo{author}{{Nagata}, S.},
  \bibinfo{author}{{Otsubo}, M.}, \bibinfo{author}{{Shimizu}, T.},
  \bibinfo{author}{{Suematsu}, Y.}, \bibinfo{author}{{Nakagiri}, M.},
  \bibinfo{author}{{Noguchi}, M.}, \bibinfo{author}{{Tarbell}, T.},
  \bibinfo{author}{{Title}, A.}, \bibinfo{author}{{Shine}, R.},
  \bibinfo{author}{{Rosenberg}, W.}, \bibinfo{author}{{Hoffmann}, C.},
  \bibinfo{author}{{Jurcevich}, B.}, \bibinfo{author}{{Kushner}, G.},
  \bibinfo{author}{{Levay}, M.}, \bibinfo{author}{{Lites}, B.},
  \bibinfo{author}{{Elmore}, D.}, \bibinfo{author}{{Matsushita}, T.},
  \bibinfo{author}{{Kawaguchi}, N.}, \bibinfo{author}{{Saito}, H.},
  \bibinfo{author}{{Mikami}, I.}, \bibinfo{author}{{Hill}, L.~D.}, \&
  \bibinfo{author}{{Owens}, J.~K.} (\bibinfo{year}{2008}).
\newblock \bibinfo{title}{{The Solar Optical Telescope for the Hinode Mission:
  An Overview}}.
\newblock {\it \bibinfo{journal}{Solar Physics}\/},  {\it
  \bibinfo{volume}{249}\/}\bibinfo{issue}{(2)}, \bibinfo{pages}{167--196}.
  \DOIprefix\doi{10.1007/s11207-008-9174-z}.
  \href{http://arxiv.org/abs/0711.1715}{\tt arXiv:0711.1715}.
\bibitem[{{Valori} et~al.(2016){Valori}, {Pariat}, {Anfinogentov}, {Chen},
  {Georgoulis}, {Guo}, {Liu}, {Moraitis}, {Thalmann} \&
  {Yang}}]{2016SSRv..201..147V}
\bibinfo{author}{{Valori}, G.}, \bibinfo{author}{{Pariat}, E.},
  \bibinfo{author}{{Anfinogentov}, S.}, \bibinfo{author}{{Chen}, F.},
  \bibinfo{author}{{Georgoulis}, M.~K.}, \bibinfo{author}{{Guo}, Y.},
  \bibinfo{author}{{Liu}, Y.}, \bibinfo{author}{{Moraitis}, K.},
  \bibinfo{author}{{Thalmann}, J.~K.}, \& \bibinfo{author}{{Yang}, S.}
  (\bibinfo{year}{2016}).
\newblock \bibinfo{title}{{Magnetic Helicity Estimations in Models and
  Observations of the Solar Magnetic Field. Part I: Finite Volume Methods}}.
\newblock {\it \bibinfo{journal}{Space Science Reviews}\/},  {\it
  \bibinfo{volume}{201}\/}\bibinfo{issue}{(1-4)}, \bibinfo{pages}{147--200}.
  \DOIprefix\doi{10.1007/s11214-016-0299-3}.
  \href{http://arxiv.org/abs/1610.02193}{\tt arXiv:1610.02193}.
\bibitem[{{van Ballegooijen} \& {Martens}(1989)}]{1989ApJ...343..971V}
\bibinfo{author}{{van Ballegooijen}, A.~A.}, \& \bibinfo{author}{{Martens},
  P.~C.~H.} (\bibinfo{year}{1989}).
\newblock \bibinfo{title}{{Formation and Eruption of Solar Prominences}}.
\newblock {\it \bibinfo{journal}{Astrophysical Journal}\/},  {\it
  \bibinfo{volume}{343}\/}, \bibinfo{pages}{971--984}.
  \DOIprefix\doi{10.1086/167766}.
\bibitem[{{Wang} et~al.(2008){Wang}, {Jing}, {Tan}, {Wiegelmann} \&
  {Kubo}}]{2008ApJ...687..658W}
\bibinfo{author}{{Wang}, H.}, \bibinfo{author}{{Jing}, J.},
  \bibinfo{author}{{Tan}, C.}, \bibinfo{author}{{Wiegelmann}, T.}, \&
  \bibinfo{author}{{Kubo}, M.} (\bibinfo{year}{2008}).
\newblock \bibinfo{title}{{Study of Magnetic Channel Structure in Active Region
  10930}}.
\newblock {\it \bibinfo{journal}{Astrophysical Journal}\/},  {\it
  \bibinfo{volume}{687}\/}\bibinfo{issue}{(1)}, \bibinfo{pages}{658--667}.
  \DOIprefix\doi{10.1086/592082}.
\bibitem[{{Wang} et~al.(2017){Wang}, {Liu}, {Ahn}, {Xu}, {Jing}, {Deng},
  {Huang}, {Liu}, {Kusano}, {Fleishman}, {Gary} \& {Cao}}]{2017NatAs...1E..85W}
\bibinfo{author}{{Wang}, H.}, \bibinfo{author}{{Liu}, C.},
  \bibinfo{author}{{Ahn}, K.}, \bibinfo{author}{{Xu}, Y.},
  \bibinfo{author}{{Jing}, J.}, \bibinfo{author}{{Deng}, N.},
  \bibinfo{author}{{Huang}, N.}, \bibinfo{author}{{Liu}, R.},
  \bibinfo{author}{{Kusano}, K.}, \bibinfo{author}{{Fleishman}, G.~D.},
  \bibinfo{author}{{Gary}, D.~E.}, \& \bibinfo{author}{{Cao}, W.}
  (\bibinfo{year}{2017}).
\newblock \bibinfo{title}{{High-resolution observations of flare precursors in
  the low solar atmosphere}}.
\newblock {\it \bibinfo{journal}{Nature Astronomy}\/},  {\it
  \bibinfo{volume}{1}\/}, \bibinfo{pages}{0085}.
  \DOIprefix\doi{10.1038/s41550-017-0085}.
  \href{http://arxiv.org/abs/1703.09866}{\tt arXiv:1703.09866}.
\bibitem[{{Wang} et~al.(1991){Wang}, {Tang}, {Zirin} \&
  {Ai}}]{1991ApJ...380..282W}
\bibinfo{author}{{Wang}, H.}, \bibinfo{author}{{Tang}, F.},
  \bibinfo{author}{{Zirin}, H.}, \& \bibinfo{author}{{Ai}, G.}
  (\bibinfo{year}{1991}).
\newblock \bibinfo{title}{{Motions, Fields, and Flares in the 1989 March Active
  Region}}.
\newblock {\it \bibinfo{journal}{Astrophysical Journal}\/},  {\it
  \bibinfo{volume}{380}\/}, \bibinfo{pages}{282--286}.
  \DOIprefix\doi{10.1086/170584}.
\bibitem[{{Wang} et~al.(2018){Wang}, {Yurchyshyn}, {Liu}, {Ahn}, {Toriumi} \&
  {Cao}}]{2018RNAAS...2....8W}
\bibinfo{author}{{Wang}, H.}, \bibinfo{author}{{Yurchyshyn}, V.},
  \bibinfo{author}{{Liu}, C.}, \bibinfo{author}{{Ahn}, K.},
  \bibinfo{author}{{Toriumi}, S.}, \& \bibinfo{author}{{Cao}, W.}
  (\bibinfo{year}{2018}).
\newblock \bibinfo{title}{{Strong Transverse Photosphere Magnetic Fields and
  Twist in Light Bridge Dividing Delta Sunspot of Active Region 12673}}.
\newblock {\it \bibinfo{journal}{Research Notes of the American Astronomical
  Society}\/},  {\it \bibinfo{volume}{2}\/}\bibinfo{issue}{(1)},
  \bibinfo{pages}{8}. \DOIprefix\doi{10.3847/2515-5172/aaa670}.
  \href{http://arxiv.org/abs/1801.02928}{\tt arXiv:1801.02928}.
\bibitem[{{Wang} et~al.(1996){Wang}, {Shi}, {Wang} \&
  {Lue}}]{1996ApJ...456..861W}
\bibinfo{author}{{Wang}, J.}, \bibinfo{author}{{Shi}, Z.},
  \bibinfo{author}{{Wang}, H.}, \& \bibinfo{author}{{Lue}, Y.}
  (\bibinfo{year}{1996}).
\newblock \bibinfo{title}{{Flares and the Magnetic Nonpotentiality}}.
\newblock {\it \bibinfo{journal}{Astrophysical Journal}\/},  {\it
  \bibinfo{volume}{456}\/}, \bibinfo{pages}{861--878}.
  \DOIprefix\doi{10.1086/176703}.
\bibitem[{{Zirin} \& {Liggett}(1987)}]{1987SoPh..113..267Z}
\bibinfo{author}{{Zirin}, H.}, \& \bibinfo{author}{{Liggett}, M.~A.}
  (\bibinfo{year}{1987}).
\newblock \bibinfo{title}{{Delta spots and great flares}}.
\newblock {\it \bibinfo{journal}{Solar Physics}\/},  {\it
  \bibinfo{volume}{113}\/}\bibinfo{issue}{(1-2)}, \bibinfo{pages}{267--283}.
  \DOIprefix\doi{10.1007/BF00147707}.
\bibitem[{{Zirin} \& {Wang}(1993{\natexlab{a}})}]{1993Natur.363..426Z}
\bibinfo{author}{{Zirin}, H.}, \& \bibinfo{author}{{Wang}, H.}
  (\bibinfo{year}{1993}{\natexlab{a}}).
\newblock \bibinfo{title}{{Narrow lanes of transverse magnetic field in
  sunspots}}.
\newblock {\it \bibinfo{journal}{Nature}\/},  {\it
  \bibinfo{volume}{363}\/}\bibinfo{issue}{(6428)}, \bibinfo{pages}{426--428}.
  \DOIprefix\doi{10.1038/363426a0}.
\bibitem[{{Zirin} \& {Wang}(1993{\natexlab{b}})}]{1993SoPh..144...37Z}
\bibinfo{author}{{Zirin}, H.}, \& \bibinfo{author}{{Wang}, H.}
  (\bibinfo{year}{1993}{\natexlab{b}}).
\newblock \bibinfo{title}{{Strong transverse fields in
  {\ensuremath{\delta}}-spots}}.
\newblock {\it \bibinfo{journal}{Solar Physics}\/},  {\it
  \bibinfo{volume}{144}\/}\bibinfo{issue}{(1)}, \bibinfo{pages}{37--43}.
  \DOIprefix\doi{10.1007/BF00667980}.

\end{thebibliography}

\end{document}